\documentstyle[aaspp4]{article}          
\lefthead{Zampieri et al.}
\righthead{Supernova fallback and the emergence of a black hole}

\begin{document}

\title{Supernova fallback and the emergence of a black hole}

\author{Luca Zampieri \altaffilmark{1},
Monica Colpi \altaffilmark{2},
Stuart L. Shapiro \altaffilmark{1,3}
and Ira Wasserman \altaffilmark{4}}

\altaffiltext{1}{Department of Physics, Loomis Laboratory of Physics,
University of Illinois at Urbana--Champaign, 1110 West Green Street, 
Urbana, IL 61801--3080}

\altaffiltext{2}{Dipartimento di Fisica, Universit\`a degli Studi di
Milano, Via Celoria 16, I--20133 Milano, Italy}

\altaffiltext{3}{Department of Astronomy and National Center for
Supercomputing Applications, University of Illinois
at Urbana--Champaign, Urbana, IL 61801}

\altaffiltext{4}{Center for Radiophysics and Space Research, Cornell
University, Ithaca, NY 14853}

\begin{abstract}

We present the first fully relativistic   
investigation of matter fallback in a supernova.
We investigate spherically symmetric supernova fallback    
using a relativistic radiation hydrodynamics Lagrangian code
that handles radiation transport in all regimes.
Our goal is to answer the fundamental question: did SN1987A   
produce a black hole and, if so, when will the hole become detectable ?
We compute the light curve, assuming that a black hole has been formed
during the explosion, and compare it with the observations.
Our preliminary calculations lack radioactive energy input
and adopt a very simple chemical composition (pure hydrogen). As a result,
our computed models cannot fit the observed data of SN1987A in detail.
Nevertheless, we can show that, during the first hours, the accretion flow
is self--regulated and the accretion luminosity stays very close
to the Eddington limit. The
light curve is completely dominated, during the
first few weeks,  by the emission of the 
stellar envelope thermal energy, and  resembles
that obtained in ``standard'' supernova theory.
Only long after hydrogen recombination takes place is there even
a chance to actually detect radiation from the accreting black hole above
the emission of the expanding envelope.
The presence of a black hole is thus not inconsistent with observations
to date.
Because of the exponential decay of the $^{44}$Ti radioactive heating
rate, the date of the emergence of the black
hole is not very sensitive to the actual parameters of the models and
turns out to be about 1000 years. The bulk of the emission then is
expected to be
%in the hard X--rays or in the $\gamma$--rays,
in the visible band,
but will be unobservable with present instrumentation.
We discuss the implications of our results in connection with the
possible emergence of a black hole in other supernovae.

\end{abstract}

\keywords{Accretion, accretion disks --- hydrodynamics --- 
methods: numerical --- radiative transfer --- relativity ---
supernovae: individual (SN1987A)}

\section{Introduction \label{sec1}}

On the 23rd of February 1987, Shelton and Jones announced the discovery of
a supernova in the Large Magellanic Cloud, SN1987A
\markcite{Kunkel et al. 1987}(Kunkel {\it et al.\/} 1987).
Historically, this was the brightest
supernova observed after that recorded by Kepler in 1604 (SN1604).
It was also the first supernova to be observed 
in every band of the electromagnetic spectrum and the first detected 
through its initial burst of neutrinos. Its relatively close distance 
($\sim$ 50 Kpc) has offered a unique opportunity to observe a supernova
in great detail and with a variety of detection techniques.
[For extensive reviews on SN1987A see e.g.
\markcite{Arnett et al. 1989}Arnett {\it et al.\/} (1989),
\markcite{McCray 1993}McCray (1993) and references therein.]
Since its appearance, SN1987A has confirmed many
aspects of the theory of type II supernovae (see e.g.
\markcite{Dar 1997}Dar 1997). As predicted,
most of the energy ($\sim 10^{53}$ erg) was released in $\sim$ 10 s
by the cooling of the protoneutron star and was radiated in form of 
neutrinos, detected by the Mont Blanc, Kamiokande, IMB and Baksan underground 
detectors. The strong shock produced by the core bounce heated and pushed
outwards the stellar envelope, which subsequently emitted its internal 
energy as photons: the integrated light emission was $\sim 10^{49}$ erg,
while the kinetic energy of the expanding shells was $\sim 10^{51}$ erg.
The light curve
can be fitted quite well by the ``standard'' type II supernova theory
(see e.g. \markcite{Woosley 1988}Woosley 1988;
\markcite{Arnett 1996}Arnett 1996). After the emergence of the shock wave,
the emission is dominated by the diffusion 
luminosity of the expanding stellar envelope during the first 30--40 days.
At $\sim 40$ days, the 
hydrogen envelope starts to recombine and a large amount of internal 
energy is released by the inward motion of the recombination front.
After most of the envelope has recombined at day $\sim 160$, the light 
curve starts to be dominated by the radioactive decay of
heavy elements synthesized during the explosion.
%by the reradiation of heat deposited by heavy elements synthesized
%during the explosion.
Some predictions of the 
``standard'' theory were not confirmed and a number of key issues 
remain to be answered, such as the nature of the 
supernova progenitor, the explosion mechanism and, perhaps most important,
the nature of the compact remnant left over after the explosion.

Stellar evolution calculations show that stars with a 
main sequence mass in the range 8--19 $M_\odot$ finish their lives
with a compact core of $\sim 1.4 M_\odot$. As shown by \markcite{Woosley
\& Weaver 1995}Woosley
\& Weaver (1995), for these stars the amount of material that falls back
toward the core in the aftermath of its collapse
is negligible, so it is likely that they give birth to neutron stars.
%during the explosion is negligible and so it is likely
%that they leave a neutrons star.
Stars with main sequence mass
larger than $M_c \sim$ 25--30 $M_\odot$ have more massive cores ($\simeq 2
M_\odot$) and undergo accretion of a significant amount of matter,
so that the remnant mass left over after the explosion is larger
than $3 M_\odot$, probably leading to the formation of a black hole.
The value of $M_c$ is somewhat uncertain and depends quite sensitively
on the mass loss prior to the explosion (reduced core mass) and on
the explosion energy (amount of fall back). The fate of stars in the
intermediate range 19--25 $M_\odot$ is far less obvious. At the end of their
evolution, they have core masses around 1.6--1.8 $M_\odot$ and a
variable amount of matter, 0.1--0.3 $M_\odot$ may fall back because
of the hydrodynamic interaction of the outgoing supernova shock with the 
expanding envelope (\markcite{Woosley \& Weaver 1995}Woosley \& Weaver
1995).
%%%; \markcite{CSW}Colpi, Shapiro \& Wasserman 1996, hereafter CSW).
Typically, these stars leave a central compact remnant of
1.7--2.1 $M_\odot$ (baryonic mass).
Then their fate depends critically on the
equation of state at nuclear matter densities that fixes the maximum
mass $M_{crit}$ above which no stable neutron star configuration can exist.
Adopting the parametrization of the nuclear matter
equation of state by
\markcite{Lattimer \& Swesty 1991}Lattimer \& Swesty (1991),
$M_{crit}$ turns out to be $M_{crit} \lesssim 2.5 M_\odot$ (gravitational 
mass). In this case, it would be unlikely that these stars could
form a black hole. However,
\markcite{Thorsson, Prakash \& Lattimer 1994}Thorsson, Prakash \& Lattimer 
(1994)
have proposed a much softer equation
of state as a consequence of $K^{-}$ condensation. They find a maximum
mass $M_{crit} \simeq 1.5 M_\odot$ (gravitational mass).
Recently, modern models of nuclear forces and accurate Monte Carlo
modeling of nucleon interactions indicate that $M_{crit} = 1.8$--2.2
$M_\odot$ (\markcite{Pandharipande 1997}Pandharipande 1997).
Based on the findings of
\markcite{Thorsson, Prakash \& Lattimer 1994}Thorsson, Prakash \& Lattimer 
(1994), \markcite{Brown \& Bethe 1994}Brown \& Bethe (1994) and
\markcite{Woosley \& Timmes 1996}Woosley \& Timmes (1996) have proposed 
different
scenarios that may lead to the formation of a black hole after the
explosion of a 19--25 $M_\odot$ star. According to
\markcite{Brown \& Bethe 1994}Brown \& Bethe
(1994), after core bounce a protoneutron star would form that could
remain stable for about $\sim 12$ s (just the time necessary to
release its internal energy in form of neutrinos) and then would
collapse to a black hole, as early suggested by
\markcite{Wilson et al. 1986}Wilson {\it et el.\/} (1986) and
\markcite{Woosley \& Weaver 1986}Woosley \& Weaver (1986).
\markcite{Woosley \& Timmes 1996}Woosley \& Timmes (1996) point out that
the formation of a black hole could be driven by the matter which
falls back in the first few hours after the explosion, as already emphasized
by \markcite{Colgate 1971}\markcite{Colgate 1988}Colgate (1971, 1988)
and \markcite{Chevalier 1989}Chevalier (1989).

Interestingly, the mass of the progenitor of SN1987A 
($M =$ 18--21 $M_\odot$) falls in the range of masses where the outcome
of core collapse is uncertain. So, from a theoretical
point of view, we do not know if a neutron star or a black hole has
been formed during the explosion. The main uncertainties
are the value of the critical mass above which no stable neutron
star can exist and the amount of matter that falls back during
the explosion. 
If there was no fallback in SN1987A, the emission from a radio pulsar 
most probably would not be detectable directly for a very 
long time, unless there are holes in the outflow. Observations also
rule out the presence of any optical pulsar. The energy deposition from
a pulsar could alter the light curve, though; observations to date limit
any energy input from a pulsar to be lower than the contribution from
radioactive decay ($\sim 10^{36}$ erg s$^{-1}$, implying $B/P^2
\sim 10^{14}{\rm G\,s^{-2}}$ for magnetic dipole emission by a pulsar
with field strength $B$ and rotation period $P$). However, high energy 
X-rays from a pulsar radiating a hard spectrum should have become observable
in a few months 
(\markcite{McCray, Shull and Sutherland 1987}McCray, Shull and Sutherland
1987) although the 1--3 keV thermal X--ray emission from the cooling of 
a neutron star would be attenuated for about 100 yr
(\markcite{Chernoff, Shapiro and Wasserman 1989}Chernoff, Shapiro and
Wasserman 1989). Such hard X-rays were never observed.

%However,
If a compact object of any kind is present,
it is probably accreting from the progenitor stellar material,
since the inner part of the ejecta have velocity smaller than
the escape velocity. During the first phases after the explosion,
the accretion rate should be as high as $10^8$ in units of the
Eddington accretion rate (\markcite{Chevalier 1989}Chevalier 1989).
On the other hand, as mentioned above,
the bolometric light curve observed to date (ten years after
the explosion) can be explained by the ``standard''
theory of Type II supernovae and does not show any feature related
to the possible presence of a central accreting compact object.
Using analytic and numerical
methods, \markcite{Chevalier 1989}Chevalier (1989) and
\markcite{Houck \& Chevalier 1991}Houck \& Chevalier (1991) have
investigated the problem of fall back onto a neutron star
and found that, for radiation pressure--limited accretion,
the luminosity emitted after 3--4 years should
approach the Eddington value, $L_{Edd} \sim 10^{38}$ erg s$^{-1}$
for gas opacity dominated by electron (Thomson) scattering.
Since at $\sim 1500$ days the bolometric 
luminosity of SN1987A was $\sim 10^{37}$ erg s$^{-1}$, they argue that 
either the accreting envelope becomes dynamically unstable or
SN1987A contains a black hole.
On the other hand,
\markcite{Chen \& Colgate 1995}Chen \& Colgate (1995) noted
that, at the typical temperature and density of the ejecta
after few years, the opacity in the accreting gas enriched of
heavy elements is much
larger than the Thomson opacity. So, they conclude that a
neutron star accreting at the actual (non--Thomson) Eddington limit may be
present. 
Furthermore, if the value of the critical mass falls in the range
1.8--2.2 $M_\odot$, as seems to be indicated by recent developments
in the nuclear many--body theory
(\markcite{Pandharipande 1997}Pandharipande 1997), it appears that
the amount of fallback is a key element
in determining whether or not a stellar black hole has formed. 
SN1987A may or may not harbor a black hole
but observations alone have not resolved this issue yet.
%It is thus of importance to establish,
%on the theoretical ground, which 
%distinctive sign a black hole or a neutron star
%(enshrouded by a hot accreting envelope)  would 
%inprint in the light curve of a successful supernova.
It is therefore important to determine what extra luminosity
an accreting central component would produce and when its
presence might be discernible.
The amount of fallback depends on the main 
parameters of the explosion while the visibility of the remnant
on the detailed hydrodynamical evolution of the flow and of the radiation 
field.

This is the first of a series of papers in which
we address a number of relevant questions: 
can we infer from the light curve whether a stellar black hole 
has formed in the aftermath of a supernova explosion ? Did 
SN1987A produce a neutron star or a black hole ?
We will consider these issues
by assuming that either a black hole or a neutron star has 
formed and then will compute the resulting light curve varying
the main parameters of the explosion.
Our focus will be to isolate the contribution from the central
compact object.

In this paper we present the first self--consistent, fully relativistic
investigation of supernova fallback in presence of a black hole.
It is the relativistic generalization of the work by \markcite{CSW}Colpi,
Shapiro \& Wasserman (1996, hereafter CSW) in which the hydrodynamical
properties of a fluid accreting onto a central 
remnant from an initially expanding cloud were first explored.
Here, we include self-consistently the transfer of radiation.
%and the study of Houck \& Chevalier (1991). 
Due to the complexity of the problem 
we will not attempt to investigate a model with realistic chemical 
composition (that will be presented in a forthcoming paper) but, instead,
we will consider a number of simpler preliminary scenarios to guide
our understanding.
We will show that, during the first hours, the accretion flow
is self--regulated and the accretion luminosity stays very close
to the Eddington limit. However, during the first few weeks the
light curve is completely dominated by the emission of the stellar
envelope internal energy, giving a light curve that resembles
that obtained in the ``standard'' supernova theory.
%Only long after that recombination takes place, is there even
%a chance of actually ``seeing'' the accreting black hole.
Although in this paper we apply our calculation specifically to SN1987A, 
our results have general validity and can be used to study the light
curve and the observational signatures of black holes in other
supernovae.

We have investigated spherically symmetric supernova fallback
using a general--relativistic, radiation hydrodynamic Lagrangian code. 
Our code can handle the transfer
of radiation from the first phases immediately after the supernova explosion
(when photons diffuse through a high temperature, expanding
cloud) to the late evolutionary stages (when the hydrogen envelope has
recombined and most of the ejecta are transparent).
The main challenge we have confronted is the enormous
dynamic range in the relevant physical timescales entering the problem.
To this end, we have implemented a multiple
timestep procedure that allows us to integrate different radial regions
at different rates (a primitive ``temporal adaptive mesh'').
This technique enables us to speed up the calculation
by almost an order of magnitude.

The plan of the paper is the following: in Section \ref{sec2} we present
the general relativistic equations of radiation hydrodynamics
used in the calculation. The finite difference form of these equations
is presented in the Appendix.
Section \ref{sec3} describes the numerical method used to solve the 
equations of radiation hydrodynamics and the boundary
conditions adopted. In Section \ref{sec4} we present in detail the multiple
timestep procedure employed in our calculation.
Section \ref{sec5} illustrates the set up of the
initial conditions. In Section \ref{sec6} we present the numerical results
for various simplified examples and test cases.
Section \ref{sec7} is devoted to the semi--analytic calculation of the 
late--time light curve. Finally Section \ref{sec8} contains a 
discussion of 
our preliminary results in connection with SN1987A and with the 
diagnosis of black holes in supernova explosions.

\section{Equations \label{sec2}}

In this section we present the equations of relativistic radiation 
hydrodynamics in spherical symmetry for a self--gravitating matter fluid
which is interacting with radiation [for a review of the derivation
of these equations see \markcite{Zampieri 1995}Zampieri 1995 and 
\markcite{Zampieri, Miller \& Turolla 1996}Zampieri, Miller \& Turolla 1996].
All the equations will be written in the frame comoving with the fluid 
flow. The 4--velocity of an element of fluid $u^{\alpha}$ will be 
evaluated in the Eulerian frame, defined as the 
reference frame at rest with respect to the background (spherically 
symmetric) metric. In this frame, the areal radius $r$ is taken to be  
the independent radial (Schwarzschild) coordinate.
We now introduce the spherically symmetric, comoving--frame line 
element
\begin{equation}
ds^2 = - a^2 dt^2 + b^2 d\mu^2 + r^2 \left( d\theta^2 +
\sin^2 \theta d\varphi^2 \right) \, ,
\label{linel}
\end{equation}
where $t$ and $\mu$ are the Lagrangian time and the comoving radial
coordinate (taken to be the rest mass contained within a comoving
spherical shell)
%$r$ is the Schwarzschild radial coordinate
and $a$ and $b$ are two functions of $t$ and $\mu$ which need to be computed
from the Einstein field equations. Here and throughout we adopt 
geometrized units and set $c = G = 1$.
Using the line element (\ref{linel}), the complete system of
radiation hydrodynamics equations in the frame comoving with the flow 
along with the Einstein field equations can be cast into the form
(\markcite{Rezzolla \& Miller 1994}Rezzolla \& Miller 1994;
\markcite{Zampieri, Miller \& Turolla 1996}Zampieri, Miller \& Turolla 1996)
\begin{eqnarray}
& & e_{,t} - h \rho_{,t} + a s_0 = 0 
%\qquad \qquad \qquad \qquad \qquad \qquad \qquad \qquad
%({\rm energy \ equation}) 
\label{energy} \\
& & u_{,t} + a \left[ {\Gamma\over b} \left( {{p_{\mu} + b s_1}\over
{\rho h}} \right) + 4\pi r \left( p + {1\over 3}w_0 + w_2
\right) + {M\over {r^2}} \right] = 0
%\ \ \ \ \
%({\rm Euler \ equation}) 
\label{euler} \\
& & {{(\rho r^2)_{,t}}\over {\rho r^2}} + a \left(
{{u_{,\mu} - 4\pi b r w_1}\over {r_{,\mu}}} \right) = 0 
%\ \ \ \ \ \qquad \qquad \qquad \qquad 
%({\rm continuity \ equation}) 
\label{continuity} \\
& & b = {1\over {4\pi r^2 \rho}}   \label{b} \\
& & {{(ah)_{,\mu}}\over {ah}} + {{h\rho_{,\mu} - e_{,\mu} + b s_1}\over 
{h\rho}} = 0  \label{a} \\
& & M_{,\mu} = 4\pi r^2 r_{,\mu} \left( e + w_0 + {u\over 
\Gamma}w_1 \right)  \label{mass}
\end{eqnarray}
where
\begin{equation}
u = \frac{r_{,t}}{a} \label{u}
\end{equation}
is the radial component of the fluid 4--velocity measured in the 
Eulerian frame, $\Gamma \equiv r_{,\mu}/b = (1 + u^2 - 2M/r)^{1/2}$,
%is the  general relativistic analogue of the Lorentz factor,
$M$ represents
the effective gravitational mass--energy (for black hole + gas + radiation)
contained within radius $r$ and $\rho$, $e$, $p$ and $h = (e + p)/\rho$
are the rest--mass density, total mass--energy density, pressure and enthalpy
of the gas flow, respectively, as measured in the comoving frame.
%The subscripts $t$ 
%and $\mu$ denote partial derivatives with respect to the corresponding 
%variables.
Here $s_0$ and $s_1$ are the radial PSTF moments of the 
source function that account for energy and momentum exchange between the fluid 
flow and the radiation field. Their actual forms depend on the radiative 
processes considered (see below).
Finally, $w_0$, $w_1$ and $w_2$ denote the first three radial PSTF 
moments of the specific intensity $I$ of the radiation field, measured in the 
comoving frame and given by
\begin{equation}
w_k = 2 \pi {{k!(2k+1)}\over {(2k+1)!!}} \int I P_k({\hat \mu}) d{\hat \mu}
\, , \end{equation}
where ${\hat \mu}$ is the cosine of the angle between the photon propagation
and radial directions and $P^k({\hat \mu})$ is the Legendre polynomial of 
order $k$. In terms of the ``classical'' moments of the specific intensity
$J$, $H$ and $K$, we have:
$w_0 = 4\pi J$, $w_1 = 4\pi H$, $w_2 = 4\pi (K - J/3)$.

The radiation hydrodynamic equations must be supplemented with the 
radiation moment equations. In spherical symmetry with the 
line--element ({\ref{linel}), the first two frequency--integrated
PSTF moments of the relativistic transfer equation can be written 
\markcite{Zampieri, Miller \& Turolla 1996}(Zampieri, Miller \& Turolla 1996)
\begin{eqnarray}
& & {1\over {b^{4/3}r^{8/3}}}
\left( w_0 b^{4/3}r^{8/3} \right)_{,t}
+ {1\over {abr^2}}(w_1 a^2 r^2)_{,\mu}
+ \left({{b_{,t}}\over b} - {{r_{,t}}\over r} \right) w_2
- a s_0 = 0 \, , \label{mom0} \\
& & {1\over {b^2 r^2}}
\left( w_1 b^2 r^2 \right)_{,t}
+ {1\over {3a^3 b}}(w_0 a^4)_{,\mu}
+ {1\over {b r^3}}(w_2 a r^3)_{,\mu}
- a s_1 = 0 \, . \label{mom1}
\end{eqnarray}
In equations (\ref{mom0}) and (\ref{mom1})
$w_0$ and $w_1$ have the dimensions of energy density.
%and $s_0$ and $s_1$ are in units of erg cm$^{-3}$ s$^{-1}$.
To close the moment equations, we employ interpolations of the
Eddington factors for a spherically symmetric, static atmosphere
in radiative equilibrium, that fit the data from
\markcite{Hummer \& Rybicki 1971}Hummer \& Rybicki (1971)
to within 30\%. The expression for the Eddington factor $f = w_2/w_0$
and the Eddington boundary factor $g = w_1(r_{out})/w_0(r_{out})$ 
are given by
\begin{eqnarray}
& & f = \left[ \frac{2}{3} - 0.59 \left( \frac{r_1}{r_{out}} \right) \right]
\frac{1}{1 + 3\tau} \, , \label{closf} \\
& & g = 1 - 0.423 \left( \frac{r_1}{r_{out}} \right) \, , \label{closg}
\end{eqnarray}
where $\tau$ is the optical depth (see below), $r_1$ is the radius at 
which $\tau = 1$ and $r_{out}$ is the outer radial boundary of the flow.
Note that $0 \leq f \leq 2/3-0.59(r_1/r_{out})$.
%If advection is not important, in the comoving frame 
%``static'' diffusion of photons takes place. In this case, the use of the 
%Hummer \& Rybicki Eddington factors to close the system of the comoving 
%frame moment equations seems reasonable. However, in the 
%present case, during the first evolutionary phase advection terms are
%important and the advection of energy by the moving fluid establishes the 
%rate of energy transport (dynamic diffusion). This fact makes the use of 
%equations (\ref{closf}) and (\ref{closg}) more questionable during this 
%phase. In addition,
%as well known, closure Eddington factors prescribed ``a priori''
%produce an error in the calculation of the moments which is at least of 
%the order of 10--20\%. As we shall see, any kind of iterative
The use of equations (\ref{closf}) and (\ref{closg}) is not exact, especially
for a moving gas. However, adopting any kind of iterative
procedure to calculate the Eddington factors precisely (see e.g.
\markcite{Shapiro 1996}Shapiro 1996) would be too expensive from
the computational point of view. Moreover, the error introduced by
equations (\ref{closf}) and (\ref{closg})
is certainly not larger than that caused by the 
use of mean opacities in the source moments (see below) and, hence,
is adequate for the present frequency--integrated calculations.

In this paper we will use very simple input physics for 
the coupled gas equations of state and the source moments.
For our simplest models we will adopt a totally ionized hydrogen gas
with constant opacity (independent of $\rho$ and $T$).
From the numerical calculations we have found that, during computation,
the maximum gas temperature is reached near the 
horizon and is not larger than $\sim 5 \times 10^8 K$.
%, so that electrons never become relativistic.
In our more detailed models we will consider
a hydrogen gas with variable degree of ionization, computed using
the Saha equation. In a forthcoming paper we will treat a more
realistic gas composition. If the electrons are nonrelativistic
($T < 10^9 K$), we have
%({\underline assumption}: H in ground state only ?)
\begin{eqnarray}
& & p = [1 + x(T)] {{\rho k_B T}\over m_p} \, , 
\label{st1} \\
& & e = \rho \left\{ 1
+ {3\over 2} \left[ 1 + x(T) \right] {{k_B T}\over {m_p}}
- [1 - x(T)] {{E_H}\over {m_p}} \right\} \, ,
\label{st2}
\end{eqnarray}
where $T$ is the gas temperature, $E_H = 13.6 \ {\rm eV}$ is the hydrogen
ionization potential and $x(T)$ is the degree of ionization, given by
\begin{equation}
x(T) = \frac{\delta}{1+\delta} \, .
\label{saha}
\end{equation}
We assume that all atoms are in their ground state.
The function $\delta$ is the positive root of the Saha quadratic equation
\markcite{Rybicki \& Lightman 1979}(Rybicki \& Lightman 1979)
\begin{equation}
\frac{\delta^2}{1+\delta} = \frac{m_H}{\rho} \left( 
\frac{2\pi m_e k_B T}{h^2} \right)^{3/2} e^{-E_H/k_B T} \, .
\label{delta2}
\end{equation}
Equation (\ref{delta2}) holds in the ``one level atom'' limit for
hydrogen, i.e. when the temperature is fairly low compared to $E_H$.
Although this is not going to be true for our calculations throughout
all the grid, however it is still true that equation (\ref{delta2}) will
give a large ionization at high temperature.
The third term inside the curly brackets in equation (\ref{st2}) accounts 
for the electrostatic potential energy of bound electrons in the neutral
hydrogen atoms. For the opacities, we adopt the interpolation formula
by \markcite{Christy 1966}Christy (1966)
which, for the range of temperatures and densities considered,
fits the hydrogen Rosseland mean opacity $k_R$ to better than 30\%:
\begin{eqnarray}
k_R = p_e \left\{ 4.85 \times 10^{-13} \frac{1}{\rho T_4} \right.
& + & \, T_4^{1/2} \left( \frac{2 \times 10^6}{T_4^4} + 2.1 \, 
T_4^6 \right)^{-1}
+ \nonumber \\
& + & \left.
\left[ 4.5 T_4^6 + T_4^{-1} \left( \frac{4 \times 10^{-3}}{T_4^4} +
\frac{2 \times 10^{-4}}{\rho^{1/4}} \right)^{-1} \right]^{-1} \right\} 
\, , \label{ross}
%& & k_P = k_R - p_e 5.4 \times 10^{-13}
%\qquad \qquad {\rm Planck \ mean} \, , \label{planck}
\end{eqnarray}
where $p_e$ is the electron pressure and $T_4 = T/10^4 K$.
%Equation (\ref{planck}) is a rough approximation.
The Planck mean opacity $k_P$ has been calculated approximately by
subtracting the Thomson scattering contribution (the first term in
the curly brackets in equation [\ref{ross}]) from the expression for $k_R$.
This choice follows from the fact that a very accurate evaluation
of $k_P$ is not critically important since the transition of the ejecta
from optically thick to thin is governed entirely by the recombination
process. Furthermore, no detailed analytic interpolation of the
Planck mean opacity in the range of temperature and density of interest
here can be found in literature and numerical interpolation from
accurate tabulated opacities would be too expensive in the present
preliminary calculation.
Using the Kirchhoff law, the first two radial moments of the source 
function $s_0$ and $s_1$ can be written
\begin{eqnarray}
& & s_0 = \rho \left( k_P B - k_0 w_0 \right) \, , \label{s0} \\
& & s_1 = - \rho k_1 w_1 \, , \label{s1}
\end{eqnarray}
where $B = a_R T^4$ ($a_R$ is the blackbody radiation constant) and
$k_0$ and $k_1$ are the absorption and flux mean opacities.
Since in the present frequency--integrated calculation the actual spectral
distribution of $w_0$ and $w_1$ is not known, we use $k_P$ and $k_R$ in 
place of $k_0$ and $k_1$, respectively. Finally, the optical depth $\tau$ 
at radius $r$ is defined by
\begin{equation}
\tau = \int_r^{\infty} k_R \rho dr \, .
\label{tau}
\end{equation}

\section{Numerical method and boundary conditions \label{sec3}}

The numerical code used to solve the equations of radiation hydrodynamics
and the radiation moment equations presented in the previous section
is based on a Lagrangian finite difference scheme with a standard Lagrangian
organization of the grid (see the Appendix; see also
\markcite{Zampieri, Miller \& Turolla 1996}Zampieri, Miller \& Turolla 1996).
The spatial and time centering of the variables ensures second
order accuracy both in space and time. 
As far as the spatial centering is concerned,
$\rho$, $B = a_R T^4$, $w_0$ and $a$ are evaluated at mid--zones,
while $r$, $M$, $u$ and $w_1$ are evaluated at zone boundaries.
To have second--order accuracy in time, $u$ and $w_1$ are both evaluated
at an intermediate time level (time--shifted).
Because of the way in which variables
are centered, the code is semi--implicit. This feature helps preserve stability
(see e.g. \markcite{Mihalas \& Mihalas 1984}Mihalas \& Mihalas 1984; 
\markcite{Shapiro 1996}Shapiro 1996). The full set of equations in 
finite difference form is written in the Appendix.

The time step is controlled by the Courant condition for stability
and by additional constraints on the fractional variation of the 
variables ($\leq$ 10\%). The Courant condition reads
\begin{equation}
a \Gamma v_c \frac{\Delta t}{\Delta r} < 1 \, ,
\end{equation}
where $v_c = (f + 1/3)^{1/2}$ is the characteristic speed of radiation 
in the Lagrangian frame (the 
fastest characteristic speed on the grid in the present problem).
The grid in Lagrangian mass $\mu$ is usually divided into
$j_{max} = 200$ zones. A constant fractional increment in grid spacing between
successive zones $\alpha$ is used. It is calculated from the equation
\begin{equation}
\mu_{j_{max}} = \mu_{j_{min}} + \sum_{j=j_{min}}^{j_{max}-1} 
\Delta\mu_{j+1/2} \, , %\alpha^{j_{max}} - R \alpha + R - 1 \, ,
\label{alpha}
\end{equation}
where $\mu_{j_{min}}$ is the inner boundary, $\mu_{j_{max}}$ is the outer 
boundary
(fixed for the conservation of the total envelope mass) and
$\Delta\mu_{j+1/2} \equiv \mu_{j+1} - \mu_j = \alpha 
\Delta\mu_{j-1/2}$.
%where $R = (\mu_{j_{max}} - \mu_{j_{min}})/\Delta\mu_{j_{min}+1/2}$.
At the beginning of the calculation $\mu_{j_{min}} = 0$.
The mass contained within the first shell $\Delta\mu_{1/2}$ is
fixed by the requirement that the radial spacing between the first two 
shells is 30\% of $r_{j_{min}}$. Equation (\ref{alpha}) can be written
as a polynomial of order
$j_{max}$ in $\alpha$ and has been solved
%(whenever a regridding occurred)
using the Newton--Raphson method.
Whenever the inner edge of the innermost zone crosses the 
inner boundary in radius $r_{in}$, it is removed from the calculation and
a regridding of all the variables is performed. During computation,
the actual inner boundary
in mass changes according to: $\mu_{j_{min}} = \mu_{j_{min}+l}$,
where $l$ is the number
of zones that have crossed $r_{in}$. The fractional increment $\alpha$ is
computed solving equation (\ref{alpha}) with the new value of $\mu_{j_{min}}$.
All the variables are then interpolated on the new grid.
The regridding procedure makes use of a 
local cubic interpolation that, for uniform spacing, reduces to a local
fourth--order Lagrangian interpolation. Performing systematic 
regriddings
is necessary both to minimize the number of grid points and to preserve 
the spatial resolution with time.

Once the finite difference representation has been introduced,
equations (\ref{euler}), (\ref{continuity}), (\ref{a}), (\ref{mass}) and 
(\ref{u}) can be solved explicitly 
(algebraically) for $u$, $\rho$, $a$, $M$ and $r$, respectively (see Appendix).
Where necessary, linear interpolation and extrapolation in time were used to 
obtain the values of quantities at suitable time levels.
Using equation (\ref{s1}) with $k_1 = k_R$,
equation (\ref{mom1}) is rewritten in the following form
\begin{equation}
{{(w_1)_{,t}}\over w_1} = - a k_R \rho
- 2 \left( {b_{,t}\over b} + {r_{,t}\over r} \right)
- {{a}\over w_1} \left[ {1\over {3 a^4 b}} (w_0 a^4)_{,\mu}
+ {1\over {a b r^3}} (f w_0 a r^3)_{,\mu}
\right]
\label{mom1c}
\end{equation}
and solved (algebraically) for $w_1$ at level $n+1/2$. The quantity
$w_2$ is expressed using the closure relation $w_2 = f w_0$
with $f$ being defined as in equation (\ref{closf}).
Again, where necessary, variables on the right hand side have been 
interpolated or extrapolated at the correct time level (see Appendix).
The energy equation (\ref{energy}) and the 0--th moment 
equation (\ref{mom0}) form a strongly coupled, nonlinear system
of equations. Using equations (\ref{s0}) with $k_0 = k_P$,
they can be cast into the form
\begin{eqnarray}
& & \epsilon_{,t} + a k_P (B - w_0) + p \left( \frac{1}{\rho} \right)_{,t}
= 0 \, , \label{energyc} \\
& & (w_0)_{,t} - a k_P \rho (B - w_0)
+ \left[ {4\over 3} \left( {b_{,t}\over b}
+ 2{r_{,t}\over r} \right) + \left( {b_{,t}\over b} - {r_{,t}\over r} 
\right) f \right] w_0 + {1\over {a b r^2}} (w_1 a^2 r^2)_{,\mu} = 0 \, ,
\label{mom0c}
\end{eqnarray}
where $\epsilon = (e - \rho)/\rho$ is the internal energy per unit
mass. They have been solved for
$B$ and $w_0$ at the new time level $n+1$ using the Newton--Raphson method 
for nonlinear systems of equations.

As far as the boundary conditions are concerned,
the time slice at constant $t$ is a characteristic direction for 
equations (\ref{a}) and (\ref{mass}). At the outer boundary we put
\begin{equation}
a = 1 \ \ \ \ \ \ \ \ \ \ \ \ \ \ \ \ \mu = \mu_{j_{max}} \, . 
\label{abc}
\end{equation}
Equation (\ref{abc}) corresponds to synchronizing the coordinate time
with the proper time of a comoving observer at the outer edge of the grid.
The inner boundary is chosen away from the origin to avoid the black hole
singularity. There we set
\begin{equation}
M = M_0 \ \ \ \ \ \ \ \ \ \ \ \ \mu = \mu_{j_{min}} \, ,
\label{mbc}
\end{equation}
where $M_0$ is the effective mass contained within the inner boundary.
At $t=0$ this is taken to be equal to the initial black hole mass $M_{bh}$.
%however, as time elapses, it increases 
%as we sum up the effective mass of all of the zones that pass 
%through the inner boundary and are removed from the calculation.
In the limit appropriate here where the infalling fluid at $\mu_{j_{min}}$
is highly supersonic with negligible thermal and radiation back pressure,
the gas behaves like dust. In this case, as time increases, we know that the
effective mass $M$ increases at the rate at which baryons (i.e. rest mass)
cross into $r_{in}$.
Once they cross, they are removed from the calculation.
The boundary conditions for the continuity and Euler
equations are
\begin{eqnarray}
& & u_{,t} + \frac{aM}{r^2} = 0 \ \ \ \ \ \ \ \ \ \ \ \ \ \ \ 
\mu = \mu_{j_{min}} \, , \label{ubc1} \\
& & u_{,t} = 0 \ \ \ \ \ \ \ \ \ \ \ \ \ \ \ \qquad \qquad
\mu = \mu_{j_{max}} \, .
\label{ubc2}
\end{eqnarray}
Equation (\ref{ubc1}) corresponds to assuming that pressure gradients
and radiative forces can be neglected at the inner boundary and that
the free--fall is achieved near to $\mu = \mu_{j_{min}}$. This turns out 
to be 
correct if the inner boundary radius $r_{in}$ is sufficiently smaller 
than the accretion radius ($r_a = GM_{bh}/c^2_s$).
%and if the flow is radiation dominated ($\Gamma = 4/3$).
%Both conditions are fulfilled in the present calculation.
During the initial evolutionary phases,
in some of our computed models radiative forces are very important and
affect the flow dynamics.
%However, the luminosity approaches the Eddington limit at around the accretion
%radius whereas, in the inner accreting region, it is sub--Eddington
%(see Section \ref{sec6}).
However, in the innermost part of the accreting region
the luminosity is always sub--Eddington (see Section \ref{sec6}).
Thus, if the inner boundary is sufficiently far in, 
at $\mu = \mu_{j_{min}}$ the gas is moving in near free--fall.
The condition at $\mu = \mu_{j_{max}}$ follows from the requirement of free 
expansion. It is correct as long as the motion of the gas near
the outer boundary is supersonic and both radiative and gravitational forces
have little influence on the outflow (see e.g. \markcite{Arnett 1980}Arnett 
1980).
%During the first evolutionary phase ($t < t_{diff}$) the diffusion luminosity
%of the expanding cloud is largely superEddington in the outer part of
%the expanding flow so that equation (\ref{ubc2}) is not precisely correct.
%However, as shown by Arnett (1980), assuming that all of the internal energy
%is converted in kinetic energy of the ejecta, the maximum fractional increment
%in the expansion velocity turns out to be $\sim$ 40\%. Since most of it occurs
%in the first doubling of the radius, equation (\ref{ubc2}) remains still valid
%for most of the evolution.
The inner boundary condition at $\mu = \mu_{j_{min}}$ for $w_1$
is a five--point Lagrangian extrapolation in $r$ along the time slice at 
constant $t$. To impose the boundary condition at
the outer edge of the grid we apply equation (\ref{mom1c}) across the
half--zone from $j_{max}-1/2$ to $j_{max}$, substituting $(w_1)_{j_{max}}
= g (w_0)_{j_{max}}$ for $(w_0)_{j_{max}}$ in the gradients with respect to 
$\mu$ (see e.g.
\markcite{Mihalas \& Mihalas 1984}Mihalas \& Mihalas 1984;
\markcite{Shapiro 1996}Shapiro 1996). The closure boundary
factor $g$ is given by equation (\ref{closg}).

As already mentioned, the evolutionary times for the present 
problem are very long in comparison with the timescales involved
in the calculation. For this reason we developed a multiple 
timestep procedure to speed up the calculation, discussed in the next section.
%that represents the more important feature of this numerical code.

\section{Multiple timestep procedure \label{sec4}}

A situation often encountered in the numerical solution of
time--dependent problems is the need to integrate the equations
over a very wide range of length scales and corresponding timescales.
Frequently, the relevant timescales vary monotonically
along the grid. In these situations, integrating
every cycle over all of the mesh with the smallest timestep
is an unwanted waste of computational
time (in particular if the number of grid points is large and/or
the evolutionary times are very long), since the evolution is unnecessarily
slow in some portions of the grid.
%The existing multiple timestep techniques, such as
%the Adaptive Mesh Refinement (AMR) method, have been designed
%to follow the development of steep features of the variables
%that appear in isolated internal regions of the integration domain.
%In AMR methods,
%a coarse grid is specified at the beginning and, during computation,
%refined subgrids are created adaptively in response to the appearance
%of features in the solution (see e.g. Berger \& Oliger 1984). 
%In principle, one could adapt such techniques to evolve
%the solution on a certain number of subgrids with different timesteps.
%However, in the present case it would not be very useful since it
%would not save a relevant amount of
%computational time. In fact, the spatial resolution of the coarse grid should
%be such to guarantee a reasonable accuracy everywhere in the integration
%domain. Then, the grid spacing and the physical conditions in the
%innermost zone automatically fix the timestep, which turns out
%to be very small (even for the coarse grid).
%%and this would automatically fix the leading timestep to be the smaller
%%one (even for the coarse grid).
We have developed an approach, that seems particularly
suitable for the present problem.
%The basic idea of our multiple timestep procedure (MTP) is to
%divide the integration domain from the beginning
%into a prescribed number of subgrids,
%each one evolving according to its own timestep.
The basic idea of our multiple timestep procedure (MTP) is to
decompose the grid into subgrids, each one evolving separately
according to its own timestep.
%No coarse grid is specified.
The subgrids with larger
timesteps are evolved with fewer steps than those with smaller ones.
The major challenge here is establishing
the communication between neighboring subgrids.
Additional complications arise if the grid is staggered in time,
so that some variables are evaluated at the full time level (time
centered) whereas other variables are evaluated at half time level
(time shifted).
Finally, if the code performs systematic regriddings
over all the grid, the evolution has to be re--synchronized
before performing any interpolation.
Our code handles all of these difficulties.

\subsection{Communication at the boundaries between neighboring subgrids}

In the following we assume that the integration domain has been divided
into $N$ subgrids.
%The boundaries of the $k$--th subgrid are labeled
%with $jmin_k$ and $jmax_k$. Clearly, $jmax_k = jmin_{k+1}$.
We assume also that the innermost zone ($k=1$) has the smallest timestep
and that the sequence of subgrids has timesteps monotonically
increasing outwards. This assumption is certainly correct in the present case.
%If the grid is staggered both in space and in time, denoting with
%$j$ and $n$ the spatial and time indices (respectively)
%the two main type of centering of a variable $A$ are: mid--zone and full time
%($A_{j+1/2}^n$), zone boundary and half time ($A_j^{n+1/2}$).
%In our numerical code
%the rest mass density $\rho$ is an example of a variable centered in
%the first way, whereas the radial component of the 4--velocity $u$
%is an example of a variable centered in the second way (see the
%Appendix). They will be used for reference in the following discussion.
In deciding the order in which to evolve the various zones, we must consider
the subgrid boundaries.
Subgrids have boundaries which are in the interior of the integration
domain and hence they will need boundary values to close the PDE's.
In the Adaptive Mesh Refinement method
a coarse grid is advanced (in time) first
and then all the subgrids are integrated to the same time level.
%One coarse grid cycle is then the basic time unit of the algorithm.
%Refined grids have boundaries which are in the interior of the integration
%domain and hence they will need boundary values not supplied with the
%differential equations. An obvious possibility (often used) is to compute
%these values by interpolation from the coarse grid (Berger \&
%Oliger 1984).
Boundary values are often computed by interpolation from the coarse grid
\markcite{Berger \& Oliger 1984}(Berger \& Oliger 1984).
By contrast, we do not evolve a coarse grid
and hence we cannot interpolate variables at the intermediate boundaries.
Instead, we evolve the various subgrids (zones) in order of increasing
timesteps (i.e. in the sequence $k=1$, $k=2$,..., $k=N$). In this way,
%we will see that
all of the subgrids, apart from the last one ($k = N$), need values
of same variables, extrapolated from earlier times,
at the outer subgrid boundary. On the other hand, boundary values
at each subgrid inner boundary can be computed by interpolation
in time from values previously calculated in the adjacent subgrid.
%from the previous (already evolved) subgrid.
Linear extrapolation and interpolation in time has been used.
%For example, consider the advance of $u$ at the outer boundary
%of the $k$-th subgrid from time level $n_k+1/2$ to $n_k+3/2$ (see Figure ...).
%In order to compute $u_{jmax_k}^{n_k+3/2}$ we need an extrapolated value
%of the density (and other variables) at point $jmax_{k}+1/2$ and time
%level $n_k+1$ (see also equation A... in the Appendix). On the other
%hand, to advance the density at the first mid--point ($jmin_{k+1}+1/2$)
%of the $(k+1)$-th subgrid from time level $n_{k+1}$ to $n_{k+1}+1$
%we need to know the velocity (and other variables) at point
%$jmin_{k+1}$ and time level $n_{k+1}+1/2$. The value of
%$u_{jmin_{k+1}}^{n_{k+1}+1/2}$ can be computed by interpolation
%using values previously
%calculated from the evolution of the $k$-th subgrid (see Figure ...
%and equation A... in the Appendix).
%Considerable care has been put in the bookkeeping of the variables,
%in particular at the subgrid boundaries.
%In this respect our procedure is very efficient
%since just one single vector has been used for each variable.

\subsection{Subgrid evolutionary algorithm}

The subgrid evolutionary algorithm
establishes the order and the conditions under which the various
subgrids have to be advanced in time. The main criterion
is to evolve the subgrids in order of increasing timestep.
%This is accomplished in the following way.
%After calculating the Courant timestep along all the integration
%domain, the code checks which of the following conditions is satisfied
%(starting from $k=2$)
Initially, the solution is evolved for a given number of cycles $n_g$ over
all of the grid with the smallest timestep.
When $n > n_g$, the subgrid evolutionary algorithm is activated.
The evolutionary time $t_k^{n_k}$ and the timestep
${\Delta t}_k^{n_k+1/2}$ of all the subgrids
are stored in a vector that is periodically updated.
The choice of the subgrid evolved at each cycle and its timestep
are decided according to the following steps:
%[1] The evolution always starts from the innermost ($k=1$) subgrid.
%[2] Before evolving any subgrid,
%the code checks whether the following inequality is satisfied
%\begin{equation}
%t_{k-1}^{n_{k-1}} - t_k^{n_k} < \frac{{\Delta t}_k^{n_k+1/2}}{\delta} 
%\qquad \qquad k \geq 2 \, ,
%\label{chk}
%\end{equation}
%where $\delta \geq 1$ is a fixed number. This test tells the code if
%the difference in the evolutionary times between subgrid ($k-1$) and $k$
%is less than $\delta^{-1}$ times the timestep of subgrid $k$.
%[3] If inequality (\ref{chk}) is satisfied, the $(k-1)$--th subgrid
%is evolved, the new timestep is computed from the Courant 
%(and any other accuracy) condition and the evolution starts again from 1.
%If it is not satisfied, subgrid $k$ is advanced
%at the time level of subgrid ($k-1$)
%and the evolution continues checking if inequality (\ref{chk}) is satisfied
%for subgrids $k$ and $k+1$ (point 2).
[1] The evolution always starts from the innermost ($k=1$) subgrid,
so that $t_1^{n_1} \geq t_2^{n_2}$. The innermost subgrid is always evolved
according to the Courant timestep, which takes its minimum value in this zone.
[2]
%The code checks whether the following inequality is satisfied
The innermost subgrid is advanced in time until the difference between
its coordinate time and the one of the neighboring subgrid does
not exceed a given threshold:
\begin{equation}
t_{1}^{n_1} - t_2^{n_2} < \frac{{\Delta t}_2^{n_2+1/2}}{\delta}
\label{chk12}
\end{equation}
where $\delta \geq 1$ is a fixed number. Inequality (\ref{chk12})
tells the code if
the difference in the evolutionary times between subgrid $k=1$ and $k=2$
is less than $\delta^{-1}$ times the timestep of subgrid $k=2$.
%[3]
%%If inequality (\ref{chk12}) is satisfied,
%If $t_{1}^{n_1} - t_2^{n_2}$ does not exceed the threshold,
%the first subgrid
%is evolved further, the new timestep is computed from the Courant 
%(and any other accuracy) condition and the evolution continues in subgrid
%$k=1$.
[3] If the difference between the coordinate times exceeds the threshold,
subgrid $k=2$ is advanced
to the time level of the first subgrid so that $t_2^{n_2} = t_1^{n_1}$.
At this point the time centered quantities of subgrids $k=1$ and $k=2$
are synchronized (but not the time shifted).
%However, in general, time shifted variables (evaluated at
%$t_k^{n_k+1/2}$) are not evaluated at the same time level.
[4] Then, the code checks whether the inequality corresponding
to (\ref{chk12}) for subgrids $k=2$ and $k=3$ is satisfied:
$t_{2}^{n_2} - t_3^{n_3} < \Delta t_3^{n_3+1/2}/\delta$. 
If the separation in time is below the threshold,
the evolution continues from the first subgrid.
[5] If it is not, subgrid $k=3$ is advanced up to the
time level of subgrid $k=2$ (which, at this stage, is also equal to
$t_1^{n_1}$). Time centered variables are then synchronized (whereas
time shifted quantities are not evaluated at the same time level).
%Time shifted variables (evaluated at
%$t_k^{n_k+1/2}$) in different subgrids
%are not evaluated at the same time level.
This procedure is repeated for all the subgrids.
Note that the factor $\delta$ has a very important role. The smaller
$\delta$ is the faster the code will evolve, since the $k$--th subgrid
will be advanced fewer times with respect to the $(k-1)$--th subgrid.
However, too small a value of $\delta$ may cause
accuracy and stability problems because of the extrapolation of the
variables at the outer subgrid boundary. By experimenting, we found
that the value $\delta \simeq 5$ turns out to be a good compromise
between computational speed and accuracy.

\subsection{Synchronization at regridding}

The innermost subgrid is regridded whenever the inner boundary
of the innermost mass shell $r_{j_{min}}$ crosses radius $r_{in}$.
%However, to preserve spatial resolution with time it is necessary
%to perform systematic regriddings over all of the integration domain.
Synchronization at regridding
represents a major problem since not all variables on the grid
are time--centered. We solve the problem in the following way. 
%As described in the previous subsection,
%the outer subgrids are advanced up to the same time level of the
%innermost subgrid whenever inequality (\ref{chk12}) (or the corresponding
%inequality for a generic subgrid $k$) is not satisfied.
Suppose that we are in the need of performing a regridding and
let ${\bar N}$ be the number of grids which are advanced up to the same time
level (see previous subsection). All the time--centered quantities of
the subgrids interior to (and including) ${\bar N}$ will be
synchronized.
%However, time shifted variables will not be synchronized.
%Suppose that, at this point,
%the innermost mass shell $r_{j_{min}}$ has crossed radius $r_{in}$.
%Then we would like to perform a regridding of all the $k \leq {\bar k}$ 
%subgrids.
As described in Section \ref{sec3}, we need to interpolate in radius
all the variables (time---shifted and time--centered). This means that also the
time--shifted variables have to be synchronized.
%If at the same time $r_{j_{min}} < r_{in}$,
This is accomplished evolving all the $k \leq {\bar N}$ subgrids together
for two cycles, i.e. forcing the timestep of all the subgrids to be equal to the
Courant timestep of the innermost region. 
In this way, both time centered and time shifted quantities in all the
subgrids with $k \leq {\bar N}$ are synchronized and a regridding
of all of the variables can be performed. If ${\bar N} = N$, all the
mesh is regridded.

\subsection{Performance}

We have tested the MTP procedure in different situations and using the
test problems which will be described in the next sections.
We have varied different parameters ($\delta$, $N$, $j_{max}$, the number of
points of each subgrid, etc.)
and the procedure turns out to be stable and accurate.
%The fractional variation between the solutions computed with
%and without the MTP procedure is less than 1\%.
The gain in computational time for a run with a grid of 200 zones
in a domain spanning 4 decades in radius is about a factor 7--8
using 4--5 subgrids. We found that, beyond a certain
number of subgrids (which depends on the specific problem
and integration domain), the gain in computational time depends mainly
on the number of points and the position of the boundary of each
subgrid and not very much on $N$.
%The reason is that the time saved in the MTP procedure
%depends mostly on the way in which the relevant timescales vary
%along the grid.
A typical run including hydrogen opacity and recombination
takes about 10 days on a Digital ALPHA station 200 4/233.

The main limit of the MTP procedure is certainly related with the
extrapolation to get boundary values at the subgrid boundaries.
This may cause problems following the evolution of shocks or any
type of sharp feature possibly present in the solution.
%(in such cases AMR methods should probably be preferred).
Problems may also originate if the flow enters a regime in which there are
very delicate balances of forces, as for example when the luminosity
is very close to the Eddington limit (within 1\%).
In such cases, it is necessary to restore the one timestep integration
for the time necessary to follow the development of sharp features
or the evolution of sensitive regimes.
However, if the prominent feature
is not crossing the boundary between two subgrids or if the delicate
regime is contained within one subgrid, the evolution can proceed
reliably using the MTP procedure.

\section{Initial conditions and relevant timescales \label{sec5}}

In a supernova explosion, fallback of material may occur
when a reverse shock forms at the interface between the 
mantle and the hydrogen envelope of the progenitor star (see e.g.
\markcite{Woosley \& Weaver 1995}Woosley \& Weaver 1995).
After the shock has propagated toward the center, the mantle is set into 
homologous expansion and matter bound to the central remnant is accreted.
A complete analysis of the fallback process should comprise
the simultaneous study of the expanding hydrogen envelope and 
of the  mantle, a region where large composition gradients develop.
Due to the complexity of the process here
we explore a simplified model that follows the radiative and dynamical 
evolution of an expanding ``cloud" of fixed  composition.
%We compute the emerging 
%luminosity over a timescale of years, neglecting 
%the sources of radioactivity that are present in 
%order to isolate the contribution from the putative black hole.
All of the emitted
energy comes from the release of heat residing in the gas originally
or is generated by compressional heating in the course of accretion onto
the black hole. Radioactive energy sources have not
been included in our calculations.
Thus, our model will not
give an accurate representation of the light curve of a supernova
which is provided, at and after the peak, by radioactivity. 
Our main purpose
is simply to calculate the contribution coming from accretion onto
the central black hole and give quantitative estimates of its importance.

At the onset of 
evolution the cloud has homogeneous density $\rho_0$ and is set into 
homologous expansion with a velocity profile $u=r/t_0,$ where $t_0$ 
denotes the expansion time scale. The maximum velocity of the ejecta
is $V_0 = u(r_{out}) = r_{out}/t_0$, where $r_{out}$ is the outer radial
boundary of the expanding cloud.
The temperature profile is taken to be equal 
to the ``radiative zero solution'' of \markcite{Arnett 1980}Arnett (1980)
\begin{equation}
T(r,0) = T_0 \left[ {{\sin(\pi x)}\over {\pi x}} \right]^{1/4} \, ,
\label{tzero}
\end{equation}
where $T_0$ is the initial temperature at the inner boundary and
$x = r/r_{out}$. 

Four parameters specify
uniquely the dynamical and thermal state of the cloud, for a fixed 
composition: the total mass $M_{cloud},$ the radius $r_{out}$,
the sound speed at the inner boundary $c_{s,0}$, and the ratio ${\tilde k}$ 
of the accretion timescale $t_{a,0}$ (defined below) to the expansion 
time $t_0.$  Five relevant timescales are involved
in this phenomenon. Using the 
electron scattering opacity (for completely ionized hydrogen)
as reference value, $k_{es} = 0.4$ cm$^2$ g$^{-1}$,
they are defined as follows:
\begin{eqnarray}
%& & t \qquad \qquad \qquad \qquad \qquad \qquad \qquad \qquad \ \
%{\rm hydrodynamical \ timescale} \\
& & t_{a,0} = {{GM_{bh}}\over c_{s,0}^3}
%\qquad \qquad \qquad \qquad \qquad \quad \
%{\rm accretion \ timescale} 
\\
& & t_0 = {{r_{out}}\over {V_0}}
%\qquad \qquad \qquad \qquad \qquad \qquad \quad
%{\rm expansion \ timescale} 
\\
& & t_{diff} = \left[{1\over {4\pi}}{k_{es}\over c}M_{cloud}\right]^{1/2}
{1\over V_0^{1/2}} 
%\qquad \qquad {\rm diffusion \ timescale \ through 
%\ the \ cloud} 
\label{tdiff} 
\\
& & t_{trans} = \left[{3\over {4\pi}} k_{es} M_{cloud}\right]^{1/2}
{1\over V_0} 
%\qquad \ {\rm timescale \ for \ transition \ 
%optically \ thick \ to \ thin} 
\label{ttrans} 
\\
& & t_{rec} = t_0 \frac{T_0}{T_{rec}}
%\qquad \qquad \qquad \qquad \qquad \qquad {\rm recombination \ timescale}
\, .
\end{eqnarray}

\noindent
$t_{a,0}$ and $t_0$ are the initial accretion and expansion timescales.
The diffusion timescale, $t_{diff}$, represents the time for the 
radiation to 
diffuse out of an homogeneous expanding cloud. It can be estimated 
considering 
$t_{diff} \sim \tau_{es}(t_{diff}) r_{out}(t_{diff})/3c$, where 
$3c/\tau_{es}(t_{diff})$ is the photon diffusion velocity and $\tau_{es}$ is 
the electron scattering optical depth. The timescale
$t_{trans}$ represents the 
time at which the 
cloud becomes transparent because of the decrease in density caused by 
the expansion; it is the time at which the photon mean free path
$\lambda(t_{trans}) = 1/k_{es}\rho(t_{trans})$
is of the order of the cloud radius $r_{out}(t_{trans})$. For 
free expansion, the density decreases as $\rho \propto t^{-3}$.
The recombination timescale $t_{rec}$ is  estimated
calculating the time at which  the gas temperature drops below  the
recombination temperature $T_{rec},$ as a consequence of 
adiabatic expansion; 
for a $\Gamma = 4/3$ polytrope, $T \propto \rho^{1/3} \propto t^{-1}$.
%The above expressions give qualitative estimates of the
%relevant times that are inherently built into the equations 
%of radiation hydrodynamics. 

We consider a series of models selected on the basis of a predetermined
hierarchy of timescales which will guide our understanding
(for the parameters and related timescales see Table \ref{tab1}).
Models 0, I, II and III deal with a cloud of fully ionized hydrogen for
which the recombination time $t_{rec} \to \infty$.
Models IVa and IVb allow for recombination: they have $T_{rec} \simeq 10^4 K$
and $t_{rec} \leq t_{diff}$.
In all models, the mass of the central black hole is taken to be
$M_{bh} = 1.5 M_\odot$.

\begin{table}
\dummytable\label{tab1}
\end{table}

The first set of models (0 and I) have been introduced mainly to test our
general relativistic radiation hydrodynamic code against previous
numerical and analytical results. The choice of the parameters
for the other models follows from the
requirement that the hierarchy of relevant timescales reflects
that for a realistic supernova model. For example, for the inferred
parameters of the hydrogen envelope of
SN1987A, we have: $t_{a,0} \ll t_0 \ll t_{rec} 
< t_{diff} \ll t_{trans}$ (see Table \ref{tab2}). The second set of runs
(models II and III) do not include recombination and for them 
$t_{a,0} \ll t_0 \ll t_{diff} \ll t_{trans} \ll t_{rec}$. In practice, 
in these models, the photons
have time to diffuse outwards before recombination takes place.
Although they do not reflect the true physics, these models are 
important to understand the role played by the central accreting 
black hole in shaping the light curve (in particular at late time).
Since they have mainly a ``pedagogical'' purpose, we decided to run 
these models with a relatively small cloud mass 
($M_{cloud} = 10^{-2} M_\odot$) and in turn small accreted mass. 
In this way $t_{diff}$ and 
$t_{trans}$ are decreased 
and, consequently, the computational time is significantly reduced.
Note that, because $t_{diff}$ and $t_{trans}$ have the same dependence
on the cloud mass, decreasing $M_{cloud}$ does not alter 
the hierarchy of timescales as long as $t_0 < t_{diff}$.
It should be noted also that in  varying $\tilde k$ (i.e. the expansion 
velocity $V_0$) the hierarchy of ``radiative timescales'' remains unchanged.
In fact, from equations (\ref{tdiff}) and (\ref{ttrans})
\begin{equation}
{t_{diff}\over t_{trans}} = {1\over \sqrt{3}} \left( {V_0\over c} \right)^{1/2} 
\, .
\end{equation}
In examining models I, II and III, we will assume that the opacities are 
given by \begin{eqnarray}
& & k_P = 10^{-2} k_{es} \left[ 1 + 10^2 e^{-(r/3r_{j_{min}+1})^2} \right]
\, , \\
& & k_R = k_{es} = 0.4 \qquad {\rm cm^2 \ g^{-1}} \, .
\label{copa}
\end{eqnarray}
The expression for $k_0$ has a maximum at $r = r_{j_{min}}$ and then goes 
rapidly to a constant value of $10^{-2} k_{es}$.
This expression has been used for convenience to keep the inner
core in LTE as long as possible in order to avoid the decrease in 
time--step caused by the inward motion of the inner boundary.

\begin{table}
\dummytable\label{tab2}
\end{table}

The last set of runs (models IVa and IVb) is a step forward to the study
of more realistic models with $t_{a,0} \ll t_0 \ll t_{rec} 
< t_{diff} \ll t_{trans}$. Table \ref{tab2} contains the
relevant parameters for SN1987A describing
the post-shock structure of hydrogen envelope and
mantle (after the reverse shock has reached the center) as 
deduced from \markcite{Arnett 1996}Arnett (1996) and
\markcite{Chevalier 1989}Chevalier (1989), respectively.
As can be seen by comparing Tables \ref{tab1} and \ref{tab2}, model IVb has 
quite realistic
initial parameters, although they do not match exactly the ones  
characterizing  the hydrogen envelope in  SN1987A.
The calculation of a SN1987A--like model would be at least 5--6 times
longer computationally. Because of the
preliminary nature of the present investigation, we have not attempted
such a calculation. 
%
%This will be done in a forthcoming paper.

\section{Numerical results \label{sec6}}

In this section we present results of the numerical calculation of the 
dynamical and thermal structure of the expanding cloud, starting from the 
initial conditions outlined in the previous section. All the light curves 
presented in the figures have been computed using $w_1$ 
calculated at the last grid point $\mu_{j_{max}}$ (which corresponds to the
outermost radius $r_{out}$). It is
\begin{equation}
L(r_{out}) = 4 \pi r_{out}^2 w_1(r_{out}) c \, .
\label{lum}
\end{equation}
In equation (\ref{lum}), $L(r_{out})$ is the luminosity measured
in the comoving frame. To obtain the luminosity seen by a distant Eulerian
observer $L_\infty$, one must transform
%the components of the flux 4--vector
to the Eulerian frame.
However, at the outer boundary $r_{out}$, general relativistic effects
can be neglected and the flow velocity $V_0$ never exceeds $0.1 c$.
Then, to calculate $L_\infty$ we can safely ignore relativistic corrections
and write
\begin{equation}
L_\infty = L(r_{out}) \, .
\label{luminf}
\end{equation}
%use the Lorentz transformation for the moments and truncate it to first
%order in $V_0$. We get
%\begin{equation}
%L_\infty = 4 \pi r_{out}^2 w_1^{s}(r_{out}) c =
%4 \pi r_{out}^2 \left[ w_1(r_{out}) + V_0 w_0(r_{out}) \right] c \simeq
%4 \pi r_{out}^2 w_1(r_{out}) c = L(r_{out}) \, ,
%\label{luminf}
%\end{equation}
%where $w_1^{s}(r_{out})$ is the flux measured in the Schwarzschild frame
%and $V_0 w_0(r_{out})$ has been neglected with respect to $w_1(r_{out})$.
%The first equality in equation (\ref{luminf}) follows from the fact that 
%the luminosity remains constant outside the expanding cloud 
%so that the luminosity seen by a distant Eulerian
%observer is equal to that measured by the Eulerian observer seating at
%$r_{out}$. This is correct if we are sufficiently far away from the central
%black hole that general relativistic effects are negligible and
%if interstellar absorption is not taken into account.

In the following we will always refer to the luminosity and accretion 
rate in units of the corresponding Eddington values for Thomson scattering
in a completely ionized hydrogen gas. It is
\begin{eqnarray}
& & l = \frac{L}{L_{Edd}} \\
& & {\dot m} = \frac{{\dot M}_0}{{\dot M}_{Edd}}
\end{eqnarray}
where $L_{Edd} = 4 \pi G M_{bh} c/k_{es} = 1.3 \times 10^{38} (M_{bh}/M_\odot)$
erg s$^{-1}$, ${\dot M}_{Edd} = L_{Edd}/c^2 = 2.3 \times 10^{-9} 
(M_{bh}/M_\odot)
M_\odot$ yr$^{-1}$
and the local accretion rate ${\dot M}_0$ has been defined by
\begin{equation}
{\dot M}_0 = 4 \pi r^2 \rho u c \, .
\label{mdot}
\end{equation}

\subsection{Model 0: polytropic accretion -- comparison with CSW}

These models have been run to test the code and to compare our
numerical results with those obtained by \markcite{CSW}CSW.
Two set of calculations were 
performed: accretion of a polytropic $\Gamma = 4/3$ gas with no radiation
(see also \markcite{Zampieri 1997}Zampieri 1997)
and accretion of a radiation dominated, completely ionized hydrogen
gas with constant opacity. In the first set of runs, we considered
two models with ${\tilde k} = 0.01$ and ${\tilde k} = 1$
and with the same value of the initial sound velocity
($c_{s,0} = 10^8$ cm s$^{-1}$). The agreement with the results of
\markcite{CSW}CSW
is good: the density profile and the accretion rate
as a function of time differ at most by $\sim 5\%$ percent. 
As already mentioned, we have tested these models changing a number
of parameters such as the number of points, the position of the
inner boundary, the number of subgrids, the position of the subgrid
boundaries and $\delta$. We found that the numerical results agree
to within a few percent. The second set of runs includes
radiation. The initial parameters are the same of the polytropic
models. The absorption opacity was taken to be sufficiently high that,
during dynamical evolution, matter and radiation were always in LTE
over most of the integration domain. Furthermore, the evolutionary time is
smaller than the diffusion timescale, so that the evolution is quasi
adiabatic. In these conditions we expect that the solution should agree
with the one obtained in the adiabatic case.
The comparison of the accretion rate versus time with the results
for a $\Gamma = 4/3$ polytrope shows agreement at the 1--2\% level.
We have also tested that, rescaling the flux mean
opacity, the flow dynamics remains unchanged. This is the expected
behavior for a radiation dominated optically thick gas,
since the radiative force $k_1 w_1 \propto (1/3b\rho) (w_0)_{,\mu}$
and the radiation energy density gradient is basically insensitive
to an overall rescaling of $k_1$.

\subsection{Model I: comparison with Arnett's analytic solution}

%During the first phases of supernova explosion, the light curve is
%dominated by the emission of the internal energy released by the shock
%wave.
If the initial accretion timescale $t_{a,0}$ is longer  than the
diffusion timescale $t_{diff}$, the thermal and radiative structure of
the envelope resembles that of a hot cloud that is cooling
because of expansion and of emission of radiation. Figure \ref{fig1} shows the
light curve for model I for which $t_{a,0} \sim t_{diff}$. The filled
squares are the computed points and the solid line gives the analytic
solution for a pure radiation pressure, constant density (and opacity) gas
given by \markcite{Arnett 1980}Arnett (1980; ``radiative zero solution'')
\begin{equation}
l = l_0 e^{-[t/t_{diff,0}+t^2/(\alpha_l t^2_{diff})]} \, ,
\label{lumarnett}
\end{equation}
where $t_{diff,0} = 3 k_{es} \rho_0 r_{out}^2/\pi^2 c$, $\alpha_l =
18/\pi^2$ and $l_0$ is the luminosity scale
%For the ``radiative zero solution'', $l_0$ is given by
%($r_s$ is the Schwarzschild radius)
\begin{equation}
l_0 \simeq {2\over 3} {{r_{out}}\over {r_s}}
{{aT_0^4}\over {\rho_0 c^2}} \, .
\label{lumarnzero}
\end{equation}
In equation (\ref{lumarnzero}) $r_s$ is the Schwarzschild radius.
Since $t_{diff,0} \gg t_{diff}$, at late times the surface luminosity falls 
as a gaussian.
%Note that the initial luminosity of the supernova is
%determined by the thermal energy per unit mass and the radius.
As can be seen from the figure, the agreement between the numerical and
analytical results is excellent.
For $t < t_{diff}$ the luminosity is constant because the decrease in the
radiation energy density gradient is exactly compensated by the increase
in both the photon mean free path and the photospheric radius. When $t$
becomes larger than $t_{diff}$, the internal energy is being exhausted
and the luminosity falls off as a gaussian.

\subsection{Model II: photon diffusion in a high temperature, expanding 
cloud with a central black hole}

For the typical parameters of a supernova with a black 
hole mass of $\sim 1.5 M_\odot$, the accretion timescale is usually 
much smaller than the diffusion timescale. The presence of the central 
black hole is ``felt'' by the expanding cloud. We have computed a model
with $M_{cloud} = 10^{-2} M_\odot$, corresponding to $t_{diff}/t_{a,0} 
\simeq 5$.  The results are shown in Figures \ref{fig2}--\ref{fig4}.
In Figure \ref{fig2}, for $t < t_{diff}$ the luminosity is roughly constant and
is still due to the emission of the cloud internal energy. The
differences with respect to the analytic solution (equation [\ref{lumarnett}])
can be explained in terms of the different hierarchy of timescales
($t_{a,0} < t_{diff}$) and of the increasing importance of the 
gas pressure. In fact, for this model the initial value of the radiation
to gas pressure ratio is not very large (in particular near to the
outer boundary where the temperature drops rapidly).
%It can be shown
%that, as time passes, the ratio $p_{gas}/p_{rad}$ tends to increase
%and becomes of order unity.
Then, equation (\ref{lumarnett}) is no longer
a good approximation to the numerical solution.
At $t \simeq t_{diff}$, the initial exponential decline
in the light curve turns into a steep power law. {\it This is the first 
signature of the presence of the central black hole}. In fact,
if no black hole were present, one would expect the light curve to fall 
off as a gaussian with time. The radiative energy emerging at this stage has
been produced by accretion during the early phases of the evolution
and transported outward diffusively. At $r_0 = 0.3 r_{a,0}$ the maximum
value of the diffusive accretion luminosity $l \simeq 0.5$ is reached at 
$t \simeq t_{a,0}$ (see Figure \ref{fig3}a) slightly after the peak in the
accretion rate (Figure \ref{fig3}b). After traveling outwards through the 
opaque cloud, this accretion energy emerges on a diffusion timescale:
this corresponds to the point, 
along the light curve, at which the analytic 
calculation fails to reproduce the actual behavior. 
Later, the cloud becomes transparent ($t \sim t_{trans}$),
and the central accreting black hole ``becomes visible''.
The steep decline of the light curve ceases and
the emergent luminosity is 
entirely produced by accretion of the gravitationally bound gas. The 
internal accreting region is still optically thick.
We note that, at times before $t_{diff}$, the heat content of 
the gas is dominated by radiation, but at times after $t_{diff}$ it
becomes dominated by the electrons, protons and hydrogen atoms. At this 
stage, the timestep of the numerical integration becomes exceedingly  small 
because the inner boundary has to be moved inward to recover the region 
where the radiation field is in LTE. In practice, this prevents us from 
studying numerically the very late--time evolution of the model. However, 
as we 
shall see later, the light curve can be equally well computed using a 
sequence of stationary models with progressively decreasing density at 
infinity (see next section).

Figure \ref{fig4} summarizes the properties of the flow at different times
along dynamical evolution. As shown in Figure \ref{fig4}a,
at the onset of expansion
the gas is outflowing with $u \propto r$. Very soon the inner low
velocity shells start to accrete
%The velocity inversion occurs
%approximately at the actual accretion radius $r_a = GM_{bh}/c^2_s$
%which is seen to move  outwards with time owing to expansion (see e.g. CSW).
%As can be seen from the figure, once the gas crosses
%the accretion radius free fall ($u \propto r^{-1/2}$) is rapidly
%achieved.
%As can be seen from the figure, once the gas inverts its motion,
and free fall ($u \propto r^{-1/2}$) is rapidly achieved.
Because the dynamical timescale in the inner part of the flow
is much smaller than that in the outer envelope, in the inner accreting region
the density profile adjusts very quickly to the
$\rho \propto r^{-3/2}$ behavior (Figure \ref{fig4}b), as in stationary flows.
In the outermost regions $\rho$ remains nearly independent of
radius and decreases with time ($\sim t^{-3}$) because of expansion
\markcite{CSW}(CSW).
The evolution of the gas temperature is closely coupled with the
properties of the radiation field (Figures \ref{fig4}c and \ref{fig4}d).
In the inner accreting region, the temperature increases inward
($T \propto r^{-5/8}$; see e.g.
\markcite{Blondin 1986}Blondin 1986 equation [17]) because of the
compressional work done by gravitational forces.
In the outer expanding region, at early times
$T$ remains roughly constant with radius,
except close to the outer boundary, where it drops significantly to
transport photons outwards diffusively. In this respect, the temperature
profile reflects the behavior of the solution of
\markcite{Arnett 1980}Arnett (1980).
As shown in Figure \ref{fig4}c, during early evolution when gas and radiation
are in LTE and $t < t_{diff}$,
$T$ drops adiabatically with time ($T \propto t^{-1}$), as expected
for a freely expanding radiation dominated gas. In Figure \ref{fig4}d, we
have plotted the luminosity profile at different times along dynamical
evolution. At $t \sim 3 t_{a,0}$,
$l$ has two maxima: the first represents the accretion luminosity
produced by the central black hole, whereas the second is due to the
emission of the internal energy stored in the cloud. 
After the initial transient phase, the accretion luminosity decreases
steadily with time.
%(because the density decreases),
The diffusive luminosity of the cloud stays almost constant up to
$t \sim t_{diff}$ and then falls off as a gaussian.
At $t \sim 500 t_{a,0}$ the cloud becomes transparent to the central
accreting region (see the curve at $t = 1034 t_{a,0}$ in Figure \ref{fig4}d).

\subsection{Model III: photon diffusion in a high temperature, {\it slowly}
expanding cloud with a central black hole}

As we already noted, the hierarchy of ``radiative'' timescales remains
unchanged as we vary the velocity of the ejecta $V_0$. So the light curve
of model III (same parameters of model II but ${\tilde k} = 0.1$; see
Table \ref{tab1}) shows  
the same basic features of model II: initial plateau due to the emission of
internal energy, a steep exponential decline turning into a power law at $t 
\sim t_{diff}$ because of the emergence of the initial transient accretion
energy, cessation of the rapid decline due to the clearing of the cloud, 
revealing the
presence of the central black hole (see Figure \ref{fig5}). However, some
differences can be noted. First, comparing Figure \ref{fig5} with Figure 2,
it appears
that the power law decline after $t = t_{diff}$ has a different slope.
This can be explained in terms of the different expansion timescale.
Since model III has an outward expansion velocity 10 times smaller than model 
II,
less work is being done by the gas and the internal energy suffers less
adiabatic degradation. Also the energy released by accretion
undergoes less degradation and can then produce a flatter tail
in the light curve after $t \sim t_{diff}$.
Second, because of the lower expansion velocity of model III, for a certain time
the outer part of the expanding cloud remains opaque whereas the central 
part is optically thin. During this phase,
the accretion luminosity produced in the inner region is partly absorbed by 
the outer optically thick region. This is the reason for the decrease in 
luminosity close to the outer boundary at $t \simeq 2400 t_{a,0}$
in Figure \ref{fig6}.
%The same problems concerning the late time evolution 
%discussed for model II apply here too.

\subsection{Model IV: clearing from expansion and recombination}

As illustrated in  Figure \ref{fig4}c, by the time that the cloud 
becomes optically thin because of expansion the temperature has already 
dropped well below $10^4 K$, the temperature at which hydrogen recombines.
In fact, in a realistic model of a supernova explosion, the clearing of the 
hydrogen envelope takes place because of recombination. For this reason
we consider a pure hydrogen gas with variable degree of ionization
in model IVa and IVb.
The ionization coefficient is computed from the Saha equation. As can be seen 
looking at the temperature and luminosity profiles of model IVa (Figures
\ref{fig8}a and \ref{fig8}b), for $t < t_{rec}$
the evolution is very similar to that of model II. At $t \sim t_{rec}$
hydrogen starts to recombine. The recombination 
process is very fast and generates a recombination wave that propagates 
rapidly through the envelope (Figure \ref{fig8}a).
Outside the recombination front the gas is optically thin, whereas inside 
the mean free path is so small that the gas is 
everywhere opaque. The recombination front almost coincides with the 
photosphere. Note the sudden rise in luminosity in Figure \ref{fig8}b after the
recombination front has propagated inward. As noted by
\markcite{Woosley 1988}Woosley (1988),
during this phase, the radiation does not have time to diffuse through the 
photosphere; the photosphere instead moves to the radiation, releasing all 
of the internal energy.
%The energy is then released {\it by} recombination, 
%but not predominantly {\it from} recombination.
The huge amount of internal energy released gives rise to a big bump in 
the light curve at $t \sim t_{rec}$ (see Figures \ref{fig7} and \ref{fig9}a).
After the maximum, the luminosity falls off abruptly (see also Figure 
\ref{fig8}b).
In contrast to models II and III, no power law 
decay can be observed after maximum. The reason is that the fast inward 
motion of the recombination front has liberated all of the internal 
energy before it could diffuse outwards. Also the initial transient accretion
energy, stored as internal energy in the gas flow, has been completely 
liberated and contributes to the big recombination bump in the light curve.
During this phase our computed light curves are similar to those
calculated by
\markcite{Woosley 1988}Woosley (1988) for stars with low envelope mass.
In this respect they resemble the light curve of a Type II--Type Ib supernova.
At late times, after that the recombination front has propagated down to
the inner accreting region, the light curve is entirely powered by
accretion onto the central black hole and the luminosity decreases
as a power law with time (see next section). In model IVb, the total mass
accreted onto the central black hole is $\sim 2 \times 10^{-5} M_\odot$.

In Figure \ref{fig8}c we follow the evolution  of the recombination front as a 
function of time (for model IVa).
%and keep track of the instantaneous position of the
%accretion radius
%(defined by evaluating the sound speed where the velocity field $u\sim 0$).
%marginally bound radius (defined as the radius below which mass shells
%are gravitationally bound).
The front clearly stalls as soon as it approaches the innermost region
where compressional heating due to accretion overcomes radiative cooling.
The density and velocity profiles are largely
unaffected by the propagation of the recombination front,
which instead influences the light cure and thermal state of the cloud.
In Figures \ref{fig9}b we plot the effective temperature, $T_{eff}$,
and the gas temperature at the photospheric radius, $T_{ph}$,
as a function of time for model IVb:
\begin{equation}
T_{eff} = \frac{L(r_{ph})}{4\pi r_{ph}^2 \sigma} \qquad \qquad \qquad
T_{ph} = T(r_{ph}) \, ,
\end{equation}
where $r_{ph}$ is the photospheric radius (i.e. the radius at which
the effective optical depth is equal to unity).
%As can be seen from Figure \ref{fig9}b, after
%an initial steep decrease, corresponding to the phase of ``adiabatic''
%expansion, the effective temperature stays roughly constant 
%($\sim 4000 K$) with time because recombination takes place and
%the photosphere starts to recede through the envelope. This is a typical
%property shown by Type II supernovae during the phase of maximum.
%When the outer envelope clears, $T_{eff}$ rises to $\sim 10^4 K$ and
%remains roughly constant with time.
As can be seen from Figure \ref{fig9}b, during the phase of ``adiabatic''
expansion, the effective temperature decreases with time to 
$\sim 4000 K$. When recombination takes place,
the photosphere starts to recede through the envelope and $T_{eff}$
rises to $\sim 10^4 K$.
From this moment on the photosphere is located in the inner accreting
region where the compressional heating is balanced by radiative
recombination.
%keeping the hydrogen partially ionized at $T \sim 10^4 K$.
%We then expect that the effective temperature will remain almost constant
%with time at around the hydrogen recombination temperature.
Furthermore, for $t > 10^3 t_{a,0}$ the gas
photospheric temperature stays very close to the effective temperature,
which suggests that the emitted continuum spectrum should be roughly a 
blackbody.
When all of the outer hydrogen envelope is 
recombined and all of its internal energy has been emitted, the fall in
luminosity ceases and the light curve starts to be dominated by the 
accretion luminosity of the gravitationally bound shells. At this point
the timestep of the numerical integration becomes very small. 
However, as we will discuss in the next section,
the asymptotic curve $l = l({\dot m})$ can be estimated using a sequence 
of stationary models with progressively decreasing density at infinity.

Consider the early behavior of model IVb (see Figures \ref{fig10}),
which has the largest initial cloud mass $M_{cloud}$.
The large amount of mass that falls back within the first few hours gives rise
to an initial (diffusion) luminosity transient that reaches the Eddington limit
(see Figure \ref{fig10}b). At this point the flow starts to be 
``self--regulated''.
As the accretion rate increases, it drives the luminosity above the Eddington
limit. Then, the radiative force eventually pushes outwards
the accreting gas (note the sign inversion of the velocity of the intermediate
shells at $t \simeq 2 t_{a,0}$ in Figure \ref{fig10}a).
The flow readjusts itself in such a way to maintain the luminosity
close to the Eddington limit. This delicate balance between radiative and
gravitational forces governs the flow dynamics and allows the gas to radiate
very close to the Eddington limit during the first evolutionary phase.
This result is in agreement with the findings of \markcite{CSW}CSW.
In their $\Gamma = 4/3$ polytropic calculation, they have shown
that in flows with $\tilde {k}\ll 1$ the pressure pushes
initially bound shells outwards, preventing them from accreting.
Since an optically thick, radiation dominated flow is dynamically equivalent
to a $\Gamma = 4/3$ polytrope, the gas pressure gradient in the
\markcite{CSW}CSW models
acts much in the same way as the radiative force in our models.

\section{Late--time light curve \label{sec7}}

\markcite{CSW}CSW showed that in polytropic flows
with ${\tilde {k}}=t_{a,0}/t_0>1$, pressure forces act as a small perturbation 
and the accreting gas behaves as a nearly collisionless fluid.
In this case the total mass accreted is (\markcite{CSW}CSW, equation [17])
\begin{equation}
\label{dustmass}
M_{dust}\simeq {8\pi\over 3}GM_{bh}\rho_0t^2_0 \, ,
\end{equation}
and the late--time evolution of the accretion rate (in units
of the Eddington rate) is equal to (\markcite{CSW}CSW, equation [29])
\begin{equation}
\label{dustmdot}
{\dot {m}}(t)\simeq {4 \pi^{2/3}\over 9}
\rho_0 \, t_0 \, c \, k_{es}
\left ({t\over t_0 }\right )^{-5/3}.
\end{equation}
In the opposite regime,  
${\tilde {k}}=t_{a,0}/t_0<1,$ pressure forces significantly affect 
the motion. After a time $t\sim 10 t_{a,0}$ the flow evolves along a sequence 
of quasi--stationary, Bondi--like states, with a time dependent
$\dot m$ determined by the slowly varying density at large distances
(\markcite{CSW}CSW, equation [45])
\begin{equation}
{\dot m}(t) \simeq  \lambda_\Gamma G M_{bh} c k_{es}
\frac{\rho_\infty}{c_{s,\infty}^3} \, ,
\label{bondimdot}
\end{equation}
where $c_{s,\infty} = (\Gamma K \rho_\infty^{\Gamma-1})^{1/2}$ is the sound
speed ($K$ and $\Gamma$ are the polytropic constant and the polytropic
index) and
$\lambda_\Gamma = 0.25[2/(5-3\Gamma)]^{(5-3\Gamma)/2(\Gamma-1)}$.
%The density at infinity was computed asking that the mass of the outer
%freely expanding shells is conserved.
For homologous expansion
$\rho_\infty = \rho_0 ( 1 + {\tilde k}t/t_{a,0} )^{-3}$.
The persistence of a Bondi--like behavior is however limited to
an interval of time between $10\,t_{a,0}<t<t_{tr}$. At late times 
$t>t_{tr},$ the flow enters the dust regime and the accretion rate
falls like ${\dot m} \propto t^{-5/3}$.
The transition time $t_{tr}$ is estimated to be 
$\sim 9{\tilde {k}}^{-3}t_{a,0}/2,$ for $\Gamma=4/3$,
and is the time when the instantaneous Bondi accretion radius
reaches a position where the fluid is unbound and flying away 
supersonically. 
%Pressure forces cause fallback to become inefficient in terms
%of the total mass accreted.
For a flow with $\tilde {k}<1$, the mass that has fallen back
%is given by
can be estimated multiplying the Bondi accretion rate ${\dot M}_{Bondi}
= 4 \pi \lambda_\Gamma \rho_0 r_{a,0}^3/t_{a,0}$ by the accretion
timescale $t_{a,0}$, which gives
\begin{equation}
\label{bondimass}
M_{Bondi}\simeq 4\pi\lambda_{\Gamma}\rho_0{[GM_{bh}]^3\over 
c_{s,0}^{6}} \, .
\end{equation}
As can be seen from the previous equation, the total accreted mass
is very sensitive to the actual value of $c_{s,0}$. $M_{Bondi}$
is only a tiny fraction $3\lambda_{\Gamma} \tilde {k}^2/2 \sim 
t^2_{a,0}/t^2_0$ of the mass accreted if the flow was collisionless
(see equation [\ref{dustmass}]). In fact, for fixed $\rho_0$ and $t_0$,
the most favorable situation (when the accreted mass is maximum)
occurs if the sound velocity is so low that pressure forces can be
neglected, i.e. if the flow is collisionless. In this regime,
$t_0$ becomes the relevant timescale for accretion and the total
accreted mass is $\sim M_{dust}$.
%Thus, $t_{a,0}\sim t_0$ is the most favorable case.
 
Models IVa and IVb have ${\tilde {k}}=1$ and ${\tilde {k}} = 0.1$ 
respectively. In IVa, the fluid is nearly collisionless and,
after time $t \simeq 10 t_{a,0}$, we find that  
$\dot {m}$ (plotted in Figure \ref{fig11}) is  well approximated
by equation (\ref{dustmdot}).
In IVb, the expansion timescale is instead larger than the accretion
time and, at intermediate times ($t_{a,0}<t<4.5\times 10^3t_{a,0}$),
the flow is expected to evolve along a sequence of quasi--steady states
characterized by $\Gamma=4/3$ and $\dot {m}\propto t^{-3/2}$
(equation [\ref{bondimdot}]).
As shown in Figure \ref{fig11}, the accretion rate decays approximately as
$t^{-3/2}$, as long as $t \lesssim 300 \,t_{a,0}$.
We then observe a slight bending toward higher rates.
%the fluid in the region of subsonic outflow becomes gas pressure dominated
%and the change in the adiabatic index causes ${\dot m}$ to increase slightly.
Later on, the accretion rate changes slope again and ${\dot m}$ starts
decaying as $t^{-5/3}$: thereafter, the flow behaves as dust.
Note that, due to the small envelope mass (for IVa) and small
${\tilde {k}}$ (for IVb), the total mass accreted is tiny in
these two models.

In Figure \ref{fig12} we have plotted ${\dot m}$ as a function of radius
at $t = 6 \times 10^3 t_{a,0}$ for model IVb.
As can be seen, in the inner region
the accretion rate is nearly independent of radius. This means that
the evolution follows a sequence of quasi--stationary states.
Indeed, the inner part of the accretion flow,
where most of the luminosity is produced,
shows the same density and temperature structure of the stationary models
of spherical accretion onto black holes
computed by \markcite{Blondin 1996}Blondin (1986),
\markcite{Park 1990}Park (1990) and
\markcite{Nobili, Turolla \& Zampieri 1991}Nobili, Turolla \& Zampieri (1991).
Therefore, we can extrapolate the late--time evolution of the
luminosity using their results. As shown in Figure \ref{fig13},
the path described along the luminosity--accretion rate plane is in fair
agreement with the
solid line, which represents the analytic expression of $l = l({\dot m})$
for stationary models derived by \markcite{Blondin 1986}Blondin (1986)
\begin{equation}
l \simeq 3 \times 10^{-7} \left(\frac{M_{bh}}{M_\odot}\right)^{-1/3}
{\dot m}^{5/6} \qquad \qquad {\dot m}\ge 10.
\label{lumblon}
\end{equation}
%Then, independent of the initial cloud density,
%the late time evolution of both models is similar
%and they tend to follow the $l = l({\dot m})$ curve of stationary models. 
Equation (\ref{lumblon}) is obtained assuming that all of the compressional
work done by the gravitational field is converted into radiation 
in the inner accreting region where $T \propto r^{-5/8}$
\markcite{Blondin 1986}(Blondin 1986). For ${\dot m} \geq 10,$
Blondin's approximation provides a satisfactory fit to the numerical value
of $l=l({\dot m})$ calculated by
\markcite{Nobili, Turolla \& Zampieri 1991}Nobili, Turolla \& Zampieri (1991).
In Figure \ref{fig14} we plot the light curves of model IVa 
and IVb computed numerically along with their asymptotic behavior.
%The late--time accretion rate is estimated using equation (\ref{dustmdot}).
For model IVa, the late--time accretion rate is estimated using equation
(\ref{dustmdot}). For model IVb, the late time ${\dot m}$ is approximately
calculated from ${\dot m}(t) = {\dot m}(t_{ref})(t/t_{ref})^{-5/3}$,
where $t_{ref} = 6.5 \times 10^3 t_{a,0}$ and ${\dot m}(t_{ref}) =
8.1 \times 10^3$ are taken from the computed model.
Figure \ref{fig14} shows that the analytic extrapolation is in fair agreement
with the points computed numerically.

In the interval $0.03 \lesssim {\dot m} \la
10$, Blondin's approximation is only approximately correct
and is invalid below $\sim 0.03$, as the flow becomes transparent to its 
own radiation. To estimate the very late--time accretion luminosity  
one can adopt (as an order of magnitude estimate)
the interpolation of the optically thin, stationary spherical accretion
models (Park 1990; Nobili, Turolla, \& Zampieri 1991)
for which 
\begin{equation}
l \propto {\dot m}^2 \qquad \qquad \qquad {\dot m \lesssim 0.03} \, .
\label{lumthin}
\end{equation}

In light of these findings, we can derive  a simple scaling 
relation for the luminosity
emitted by the black hole soon after the recombination front has
propagated down to the inner accreting region. 
%This is the instant of ``maximum visibility" of the 
%black hole since the accretion rate
%then slowly declines with time, as illustrated in Figure 7 and 9a.
The scaling involves the main physical parameters of the flow at the
onset of evolution.
If the fluid is collisionless (hypothesis appropriate for model IVa) 
we can estimate the accretion rate according to equation (\ref{dustmdot}) 
\begin{equation}
{\dot {m}}(t_{rec})\simeq {4 \pi^{2/3}\over 9}
\rho_0 \, t_0 \, c \, k_{es}
\left ({t_{rec}\over t_0}\right )^{-5/3} \, .
\label{dustrec}
\end{equation}
Adopting equation (\ref{lumblon}) as scaling relation between
$l$ and $\dot {m}$, we find a luminosity at $t_{rec}$
\begin{equation}
L_{rec}\sim 1.2 \times 10^{40} 
\left (\rho_0 t_0 \right )^{5/6}
\left ({t_{rec}\over t_0}\right )^{-25/18}\,{\rm {erg\,\,s^{-1}}} \, .
\label{lumdrec} 
\end{equation} 
Using as reference values those of Table \ref{tab1} and a value of  
$T_{rec}\sim 10^4$K, for model IVa we estimate a luminosity $L_{rec}\sim
1.4\times 10^{35}\rm {erg\,\,s^{-1}}$ which
is in close agreement we the numerical results.
Equation (\ref{lumdrec}) thus provides an approximate expression for
the accretion luminosity soon after clearing by recombination.
$\,L_{rec}$ is a function of $\rho_0$, $T_0$ and $t_0$ at the onset of
the explosion. Subsequently,
the luminosity will decline with time as $t^{-25/18}$ (from equations
[\ref{dustmdot}] and [\ref{lumblon}]).

If the flow is radiation pressure dominated (model IVb)  
we can instead estimate the accretion rate at $t\simeq t_{rec}$  from  
equation (\ref{bondimdot}). Thus, we have 
\begin{equation}
{\dot {m}(t_{rec})}\simeq \lambda_{4/3}
GM_{bh} c\,k_{es} {\rho_0\over c_{s,0}^3}
\left ({t_{rec}\over t_0}\right )^{-3/2}
\label{mdotrec}
\end{equation}
and the luminosity turns out to be
\begin{equation}
L_{rec}\sim 5.4
\label{lumbondi}
\times 10^{23} L_E 
\left ({\rho_0\over c_{s,0}^3}\right )^{5/6}
\left ({t_{rec}\over t_0}\right )^{-5/4} \, .
\end{equation}
For model IVb $L_{rec}\sim 10^{35} \rm{erg\,\,s^{-1}}$.
The above relation can be applied to our numerical models since the
density evolution is seen to preserve a self-similar character, 
even during the propagation of the recombination front:
inside the accretion region, 
$\rho(r,t)\simeq \rho_0(t/t_0)^{-3}[r/r_{a}(t)]^{-3/2}$,
where $r_a(t)\simeq r_{a,0}(t/t_0)$ denotes the current value of 
the accretion radius.
Equation (\ref{lumbondi}) fixes the ``level" of the luminosity
at the moment of clearing of the envelope. Its time evolution
will then depend on the nature of the flow after $t_{rec}$.
At late times we expect the transition to dust giving a  
luminosity $L\propto t^{-25/18}$ (as long as ${\dot m} \gtrsim 10$).

In the next section we will use this scaling to give a first estimate of the 
time at which the bolometric light curve of SN1987A becomes dominated by
the accreting black hole.

\section{Discussion and conclusions \label{sec8}}

%Because of the lack of the radioactive energy input and the very simple
%chemical composition, our computed models
%cannot fit the observed data of SN1987A. Producing such a fit
%is not the purpose of the present paper (for the complications
%in fitting the light curve of SN1987A see e.g.
%\markcite{Woosley 1988}Woosley 1988).
%In particular, the actual shape and amplitude of the peak cannot be reproduced.
%Among the computed models, the one with initial parameters most similar to
%those inferred for SN1987A is model IVb. In particular, the radius $r_{out}$
%and the sound velocity $c_{s,0}$ (which is strictly related to the
%explosion energy per unit mass) are very close to the realistic case.
%On the other hand, the initial cloud mass is about 10--15 times smaller and
%the initial expansion velocity $V_0$ is about 4 times larger.

The computed light curves of model IVa and IVb show that, 
prior to recombination, the release of internal energy
from the expanding stellar envelope dominates over accretion.
Although during the first few hours the diffusive
accretion luminosity can be very close to the Eddington limit,
the diffusive and advection luminosities of the hot
outer layers are largely superEddington. As shown by model IVb,
during recombination the computed light curve 
shows the typical behavior observed in 
Type II--Ib supernovae with low envelope mass.
The luminosity at peak is comparable to that of SN1987A 
because the initial sound velocity of the hydrogen envelope
is similar (see Tables \ref{tab1} and \ref{tab2}).
The high expansion velocity and low envelope mass of model IVb
reduce all of the relevant timescales (see Table \ref{tab1}) so that 
recombination
takes place at about 10 days after the onset of the explosion, instead of
the 40--50 days inferred for SN1987A (see e.g.
\markcite{Woosley 1988}Woosley 1988). Nevertheless, our numerical results
clearly indicate that no distinct signature in the light curve is found  
that could reveal 
the accreting black hole during this initial evolutionary phase.
Only after the sharp decrease in luminosity due the clearing of the 
envelope by recombination,
the accreting black hole starts to power 
the bolometric light curve of the computed models and $l$ decreases
as $t^{-25/18}$ (see equations [\ref{dustmdot}] and [\ref{lumblon}]). 

Slightly after maximum, SN1987A
was powered by the radioactive decay of $^{56}$Co and $^{57}$Co and,
at present, its light curve is
%declining at 1 magnitude per 1000 days or less
consistent with emission from the decay of $^{44}$Ti
(\markcite{Suntzeff 1997}Suntzeff 1997; see  Figure \ref{fig14}). 
Can we discern the luminosity emitted by the
accreting  black hole above the contribution resulting from $^{44}$Ti ?
As already noted, the accretion history is very sensitive to the
value of the sound speed (see equation [\ref{bondimass}]). Depending
on the actual value of $c_{s,0}$ in the inner part of the ejecta
surrounding the compact remnant (the mantle), the accretion rate
and, in particular, the total accreted mass vary significantly.
An upper limit to the accretion luminosity can be inferred taking
the initial parameters of the mantle quoted by
\markcite{Chevalier 1989}Chevalier (1989; see
Table \ref{tab2}). In this case, $c_{s,0} \simeq 3 \times 10^7$ cm s$^{-1}$
and $t_{a,0} \sim t_0$, giving substantial fallback.
The mass accretion rate can be computed using equation (\ref{dustmdot}) and
the luminosity according to equation (\ref{lumblon}). 
Thus, we have
%an upper bound for the accretion luminosity
\begin{equation}
\label{upper}
l \simeq 3\times 10^{-7}(\rho_0 \, t_0 \, c \, k_{es})^{5/6}\left 
({t\over t_0}\right)^{-25/18}.
\end{equation}
%Chevalier (1989) estimated $\rho_0\, t_0^3\approx 10^9$ cgs and $t_0\approx 
%7000$ seconds (Table \ref{tab2}), for which equation (\ref{upper}) implies
%a maximum accretion luminosity
Following \markcite{Chevalier 1989}Chevalier (1989), we adopt $\rho_0\, 
t_0^3\approx 10^9$ cgs and $t_0\approx 7000$ seconds, yielding a
maximum accretion luminosity
\begin{equation}
\label{lumfall}
l \simeq {8\times 10^{-3}\over (M_{bh}/M_\odot)^{1/3}[t({\rm 
years})]^{25/18}} \, .
\end{equation}
Equation (\ref{lumfall}) implies $L\simeq 5\times 10^{34}{\rm erg\,s^{-1}}$ 
after $t\approx 10$ years, which is well below the present day
bolometric luminosity of the remnant ($\sim 10^{36}{\rm erg\,s^{-1}}$;
Suntzeff 1997) and also smaller than the luminosity estimated to result
from radioactive decay. Thus, there is no observation that rules out the
possibility that a black hole resides inside the SN1987A remnant.
%At present the upper limit on the luminosity is estimated to be
%\begin{equation}
%\label{sn87a}
%L\simeq 5\times 10^{34}
%\,\,\rm {erg\,s^{-1}}.
%\end{equation}
%Then, the present bolometric luminosity of SN1987A cannot be accounted
%for by accretion onto a putative central black hole.
%The presence of a black hole is thus not inconsistent with observations
%to date.
At very late times, when $\dot m \la
0.03$, the corresponding upper bound would be
\begin{equation}
\label{late}
l \simeq 2\times 10^{-8}(\rho_0 \, t_0 \, c \, k_{es})^2
\left ({t\over t_0}\right )^{-10/3},
\end{equation}
or, using Table \ref{tab2},
\begin{equation}
\label{afterti}
l = {7\times 10^{-8}\over [t({\rm 10^3 years})]^{10/3}}. 
\end{equation}
In Figure \ref{fig14} we plot the late--time light curve
of SN1987A calculated using equations (\ref{lumfall}) and (\ref{afterti}).
%: the accretion luminosity crosses the extrapolated bolometric light 
%curve of SN1987A at $t \sim 900$ years.
As radioactive decay plummets at 
around %100,000-1,000,000 days, or 
270--2700 years, it is clear that
the black hole would only appear after about 900 years irrespective of the
detailed numbers. 
After this time has elapsed, the luminosity of the remnant would be
$\sim 10^{32}{\rm erg\,s^{-1}}$, too dim to be detectable using present
technology, and possibly even a challenge for our distant descendants, as
it would be hard to distinguish from the multitude of low mass
stars crowding its field in the Large Magellanic Cloud.

Since we have carried out a frequency--integrated calculation,
we can only estimate the emission properties
from the effective temperature and the photospheric gas temperature.
During the optically thick accretion phase (when $\dot {m}>0.03$),
the numerical results show that $T_{eff}$ and 
$T_{ph}$ are very close and have a value  $\sim$ 8000--10000 $K$.
This corresponds to a mean photon
energy in the few eV range, for which the bulk of the emission is expected
in the visible band. Since $T_{eff}$ and $T_{ph}$ are close together
the continuum spectrum should be roughly a blackbody. This is not surprising
because the emission processes are thermal and the gas in the inner accreting
region is optically thick.
%%LUCA//IRA at present....what would be the effective temperature ?????
%%%% please consider what is written below!!!!!!!!!!!!!!
%However, it should be noted that some deviations
%from a Planckian should be certainly expected because, for free--free 
%processes, the photospheric radius varies with photon frequency.
Only at very late times ($t \gtrsim 2500$ yr), the accretion flow becomes
optically thin and the gas temperature reaches very high values
($10^8$--$10^9 K$; this regime has not been treated in the present numerical
computation).
%So, at the time of emergence, the gas will be emitting optically thin,
%high temperature thermal radiation and, hence, the continuum spectrum
%will be peaked in the hard X--rays or in the $\gamma$--rays.
So, at the time of emergence, the inner part of the accretion flow
will be still optically thick and will be emitting roughly a black body
spectrum peaked in the visible band.

%Despite the limitations of our preliminary analysis, 
%we have determined an upper limit on the current value of the 
%accretion luminosity emitted by the putative black  
%hole of $ 5\times 10^{34}$ erg s$^{-1},$ in SN1987A.  
%This luminosity is emitted at a characteristic
%temperature $\simeq 10000$ K.
%Though well below the contribution by radioactive elements, 
%this radiation is emitted with a continuum spectrum contrary to
%the emission from radioactivity that comes in lines.
%Using indirectly Crotts, Kunkel, \& Heathcote (1995)  
%upper limit on the evidence of a companion main-sequence star,
%Chevalier (1996) deduced a bound on the luminosity in the continuum of
%$\sim 2.4\times 10^{35}$ erg s$^{-1}$ (for a black body at $3000$ K).
%Our inferred upper limit for the luminosity of the
%accreting black hole is higher by a factor of 10 than Chevalier's
%estimate of $4\times 10^{33},$ and more
%accurate spectral information is necessary if one whishes
%to infer the presence of the black hole in SN1987A.
%The time of emergence of the black
%hole is not very sensitive to the actual parameters of the models and
%turns out to be $t \sim$ 1000 years;
%thus supernova remnants laking of an internal energy source might hide
%a black hole. 
 
A number of interesting issues can be addressed within the framework
presented in this paper. Can the presence of a stellar black
hole in SN1987A be deduced from the excess accretion luminosity it produces ?
What is the observational signature of the explosion of more massive
progenitors (those with $M\sim 30 M_{\odot}$) on the light curve of the
remnant ? Woosley, \& Timmes (1996) suggested that a distinguishing
signature would be a bright plateau that steeply falls to very low or zero
luminosity, similarly to what found in our models.
%As an example, we verified that doubling the explosion timescale $t_0$
%and outer radius $r_{out}$ relative to the values deduced for SN1987A, but
%keeping the entropy the same, leads to an accretion luminosity that is
%still larger than the emission resulting from radioactive decay after
%$X$ years.'
 
A number of  assumptions limit the applicability of our results, amongst,  
the hypothesis of spherical symmetry and the simple
chemical composition considered. As a result of
an asymmetry in the explosion mechanism, black holes may come to birth 
with significant intrinsic velocities (a few $\times \, 100$ km $^{-1}$).
This is certainly observed for pulsars
whose mean birth speed is around 250--300 km s$^{-1}$ \markcite{Hansen
\& Phinney 1997}(Hansen \& Phinney 1997;
\markcite{Cordes \& Chernoff 1997}Cordes \& Chernoff 1997).
%However, as noted by Chevalier (1989),
%since the initial sound speed of the ejecta is of the order of $10^3$
%km $^{-1}$, the initial accretion rate would not be significantly modified
%by the intrinsic black hole velocity.
However, since the initial sound speed of the ejecta can be of the order of 
$10^3$ km s$^{-1}$, the accretion will proceed essentially subsonically.
The central compact remnant
tends to comove with gas expanding at its velocity and, if the ambient
density is uniform, the conditions are independent of position within
the core, as in the cosmological Hubble flow. Thus, even a relatively large
intrinsic velocity should not strongly modify the results presented here.
Deviations from spherical symmetry can arise if infalling matter has even
a small amount of specific angular momentum, which presents a barrier to
spherical infall (\markcite{Chevalier 1996}Chevalier 1996).
%If an  advection--dominated disk forms, accretion would proceed 
%without neutrino losses; the luminosity could be 
%higher than in the spherical case (due to the higher efficiency) rising the 
%chance of seeing the black hole.
In the case of SN1987A, the fact that the observed neutrino 
burst is consistent with implosion models without rotation
\markcite{Burrows 1988}(Burrows 1988) seems to indicate that
angular momentum of matter initially close to the black hole was not
very important (see also \markcite{Chevalier 1996}Chevalier 1996).
In this case, the accretion should 
have proceeded almost radially, although mixing of outer mantle material
with significant angular momentum into the inner region could modify
our picture.

A more realistic model with shell--like chemical composition and heating
due to radioactive decay will enable us  to address 
more quantitatively the problem of the
visibility of black holes in otherwise successful supernova.
This analysis will also set the basis for studying fallback onto  
a nascent neutron star in order to verify whether major
outflows establish that reverse the process, during the decline
of the supernova light curve.
%In future we will proceed along these lines.

\acknowledgments

This work was supported in part by NSF grants AST 93--15133,
AST 96--18524 and NASA grants NAG--5--2925 and NAG--5--3420
at the University of Illinois at Urbana--Champaign and by
NSF grant 93-15375 and NASA grant NAG--5-3097 at Cornell University.
Monica Colpi would like to thank the Department of Physics of
the University of Illinois at Urbana--Champaign for its hospitality
during part of this work. Luca Zampieri would like to thank Luciano Rezzolla
for carefully reading and suggesting improvements of Section \ref{sec4}.

%>>>>>>>>>>>>>>>>>>>>>>>>>>>>>>>>>>>>>>>>>>>>>>>>>>>>>>>>>>>>>>>>>>>>>>>>>>>

\appendix
\section{Finite difference equations}

The full set of finite difference equations of relativistic hydrodynamics
plus the radiation moment equations are summarized in this appendix.
The indexes $j$ and $n$ will be used to denote spatial and time dependence,
respectively. As discussed in the main text, $\rho$, $B$, $w_0$ and $a$ are 
evaluated
at mid--zones ($j+1/2$), while $r$, $M$, $u$ and $w_1$
are evaluated at zone boundary ($j$).
Mid--zone variables can be easily computed from zone boundary variables
as: $A_{j-1/2} = (A_j + A_{j-1})/2$, where $A$ is any variable.
To compute zone boundary quantities from mid--zone quantities, it is necessary
to take into account for the nonuniform spacing of the grid. Using linear
interpolation, it is
\begin{eqnarray}
A_j = (\alpha A_{j-1/2} + A_{j+1/2} )/(1+\alpha) \, ,
\label{fdint}
\end{eqnarray}
where $\alpha$ is the fractional increment in grid spacing between
successive zones (see Section \ref{sec3}).
As far as time centering is concerned, $u$ and $w_1$
are evaluated at an intermediate time level ($n+1/2$; time--shifted),
while all the other variables are centered at the full time level ($n$).
Linear interpolation/extrapolation in time has been used when necessary.

Initial values for all the variables are specified on the grid according
to what discussed in Section \ref{sec5}. Before starting the evolution,
the radial component of the flow velocity $u$ and the radiative flux $w_1$
are advanced at time level $t^{1/2}$ solving equations (\ref{euler})
and (\ref{mom1c}) with a forward time integration scheme.

The code starts to evolve first equation (\ref{u}). Solving for the Eulerian
radius at the new time level, we obtain ($j_{min} \leq i \leq j_{max}$)
\begin{equation}
r_j^{n+1} = r_j^n + \Delta t^{n+1/2} \, a_j^{n+1/2} u_j^{n+1/2} \, ,
\label{fd1}
\end{equation}
where $\Delta t^{n+1/2}$ is the timestep between level $n$ and $n+1$ and
$a_j^{n+1/2}$ is interpolated at the zone boundary in space and extrapolated
at time level $n+1/2$. Then, the time evolution of the matter
density is computed. Equation (\ref{continuity}) can be formally integrated
in time and solved for $\rho_{j-1/2}^{n+1}$ using the Crank--Nicholson
operator (see e.g. \markcite{May \& White 1967}May \& White 1967;
$j_{min}+1 \leq j \leq j_{max}$)
\begin{equation}
\rho_{j-1/2}^{n+1} = \rho_{j-1/2}^n \, 
\frac{(r^2)_{j-1/2}^n}{(r^2)_{j-1/2}^{n+1}}
\, \frac{1-F_\rho/2}{1+F_\rho/2} \, ,
\label{fd2}
\end{equation}
where 
\begin{equation}
F_\rho = \Delta t^{n+1/2}
a_{j-1/2}^{n+1/2} \left[ \frac{u_j^{n+1/2} - u_{j-1}^{n+1/2}}{r_j^{n+1/2}
- r_{j-1}^{n+1/2}} -
2\pi \frac{r_j^{n+1/2}(w_1)_j^{n+1/2} + r_{j-1}^{n+1/2}(w_1)_{j-1}^{n+1/2}}
{\Gamma_{j-1/2}^{n+1/2}} \right] \, .
\label{fd2f}
\end{equation}
In equation (\ref{fd2}), $(r^2)_{j-1/2}^{n+1}$ has been computed in order
to give the correct total volume of shell $j-1/2$
(\markcite{May \& White 1967}May \& White 1967).
To derive equation (\ref{fd2}) and (\ref{fd2f}), we used equation (\ref{b})
and the two following relations, valid at constant Lagrangian time $t$:
\begin{eqnarray}
& & r_{,\mu} = b\Gamma \\
& & \frac{\partial}{\partial \mu} = r_{,\mu} \frac{\partial}{\partial r} \, .
\end{eqnarray}
In equation (\ref{fd2f}), $a_{j-1/2}^{n+1/2}$ and $\Gamma_{j-1/2}^{n+1/2}$
have been extrapolated at time level $n+1/2$.

After calculating the optical depth
$\tau_j^{n+1}$ and the Eddington factor $f_j^{n+1}$ with the advanced value
of the matter density, the energy and the 0--th moment equations
(equations [\ref{energyc}] and [\ref{mom0c}]) are solved for $B_{j-1/2}^{n+1}$ 
and
$(w_0)_{j-1/2}^{n+1}$ ($j_{min}+1 \leq j \leq j_{max}$)
%using the Newton--Raphson method for non--linear systems of equations
\begin{eqnarray}
\frac{\epsilon_{j-1/2}^{n+1} - \epsilon_{j-1/2}^n}{\Delta t^{n+1/2}}
& + & a_{j-1/2}^{n+1/2} (k_P)_{j-1/2}^{n+1/2}
\left[ B_{j-1/2}^{n+1/2} - (w_0)_{j-1/2}^{n+1/2} \right] + \nonumber \\
& + & \frac{p_{j-1/2}^{n+1/2}}{\Delta t^{n+1/2}}
\left( \frac{1}{\rho_{j-1/2}^{n+1}} - \frac {1}{\rho_{j-1/2}^n} \right) = 0
\label{fd3}
\end{eqnarray}
\begin{eqnarray}
\frac{(w_0)_{j-1/2}^{n+1} - (w_0)_{j-1/2}^n}{\Delta t^{n+1/2}}
& - & a_{j-1/2}^{n+1/2} (k_P)_{j-1/2}^{n+1/2} \rho_{j-1/2}^{n+1/2}
\left[ B_{j-1/2}^{n+1/2} - (w_0)_{j-1/2}^{n+1/2} \right] + \nonumber \\
& + & (w_0)_{j-1/2}^{n+1/2} a_{j-1/2}^{n+1/2} \left[ \left( \frac{4}{3}
+ f_{j-1/2}^{n+1/2} \right) \frac{1}{(r^2)_{j-1/2}^{n+1/2}} \times \right. 
\nonumber \\
& & \left. \times \frac{ u_j^{n+1/2}(r_j^{n+1/2})^2 - 
u_{j-1}^{n+1/2}(r_{j-1}^{n+1/2})^2 }
{ r_j^{n+1/2} - r_{j-1}^{n+1/2} } - 3 \left(f \frac{u}{r} 
\right)_{j-1/2}^{n+1/2}
\right] + \nonumber \\
& + & \left( \frac{\Gamma}{ar^2} \right)_{j-1/2}^{n+1/2}
\left[ \frac{ (w_1)_j (a_j)^2 (r_j)^2
- (w_1)_{j-1} (a_{j-1})^2 (r_{j-1})^2 }
{ r_j - r_{j-1} } \right]^{n+1/2} - \nonumber \\
& - & \left[ \frac{4\pi r a}{\Gamma} \left( \frac{4}{3} + f \right) w_0 
w_1 \right] _{j-1/2}^{n+1/2} \, ,
\label{fd4}
\end{eqnarray}
where $B_{j-1/2}^{n+1/2} = (B_{j-1/2}^n + B_{j-1/2}^{n+1})/2$ and
$(w_0)_{j-1/2}^{n+1/2} = [(w_0)_{j-1/2}^n + (w_0)_{j-1/2}^{n+1}]/2$.
Since $\epsilon$, $p$ and $k_P$ are in general rather complicated functions
of $\rho$ and $B$ (see e.g. equations [\ref{st1}], [\ref{st2}]),
equations (\ref{fd3}) and (\ref{fd4}) form a highly non--linear system
that has been solved by means of the Newton--Raphson method. To write the 0--th
moment equation in finite difference form,
$r_{,t}$ has been substituted with $au$ (equation [\ref{u}]) and the term
$b_{,t}/b = - (\rho r^2)_{,t}/\rho r^2$ has been expressed using the continuity
equation (\ref{continuity}).

%The metric coefficient $a$ is calculated from equation (\ref{a})
The time slice at constant $t$ is a characteristic direction for equation
(\ref{a}) so that it can be formally integrated along the grid
at the new time level $t^{n+1}$. Using the Leith--Hardy operator
to approximate the exponential $\exp(-F_a) =
\exp \{-\int [(e_{,\mu} - h\rho_{,\mu} - bs_1)/h\rho] d\mu \}$ in the interval
$[\mu_{j-1/2},\mu_{j+1/2}]$
(see \markcite{May \& White 1967}May \& White 1967), it reads
($j_{min} + 1 \leq j \leq j_{max}$)
\begin{equation}
a_{j-1/2}^{n+1} = \frac{(ah)_{j+1/2}^{n+1}}{h_{j-1/2}^{n+1}}
\frac{1}{1+F_a+F_a^2/2} \, ,
\label{fd5}
\end{equation}
where 
\begin{eqnarray}
F_a = \left\{ \frac{1}{(h\rho)_j}
\left[ \rho_j \left( \epsilon_{j+1/2} - \epsilon_{j-1/2} \right)
- \frac{p_j}{\rho_j} \left( \rho_{j+1/2} - \rho_{j-1/2} \right)
- \frac{ \mu_{j+1/2} - \mu_{j-1/2} }{ 4\pi r_j^2 \rho_j } (s_1)_j \right]
\right\}^{n+1} \, .
\label{fd5f}
\end{eqnarray}
In equation (\ref{fd5f}), $\rho_j$, $p_j$ and $(h\rho)_j$
are linearly interpolated at the zone boundaries.
%taking into account for the non uniform spacing of the grid
%\begin{equation}
%A_j = (\alpha A_{j-1/2} + A_{j+1/2} )/(1+\alpha) \, ,
%\label{fdint}
%\end{equation}
%where $A$ is any of the quantity $\rho$, $h$, $p$.
The difference $\mu_{j+1/2} - \mu_{j-1/2}$ can be
written as $(1+\alpha) \Delta \mu_{j-1/2}/2$. Finally, the value of $w_1$ in
$s_1$ has been linearly extrapolated at the new time level $t^{n+1}$.
The boundary condition at $\mu = \mu_{j_{max}}$ fixes $a_{j_{max}} = 1$
(equation [\ref{abc}]).

After that the new timestep $\Delta t^{n+3/2}$ has been computed,
the radial component of the fluid
4--velocity is calculated from the Euler equation (\ref{euler})
($j_{min}+1 \leq j \leq j_{max}-1$)
\begin{eqnarray}
u_j^{n+3/2} = u_j^{n+1/2} & - & \Delta t^{n+1} a_j^{n+1}
\left\{ \frac{\Gamma_j}{h_j} \left[ \frac{8\pi r_j^2}{1+\alpha}
\left( \frac{p_{j+1/2} - p_{j-1/2}}{\Delta \mu_{j-1/2}} \right) +
\frac{(s_1)_j}{\rho_j} \right] + \right. \nonumber \\
& + & \left.
4\pi r_j \left[ p_j + \left(\frac{1}{3} + f_j \right) (w_0)_j \right]
+ \frac{M_j}{r_j^2} \right\}^{n+1} \, ,
\label{fd6}
\end{eqnarray}
where $\Delta t^{n+1}$ is defined below (equation [\ref{timestep}]) and
$\Gamma_j$, $h_j$, $p_j$, $\rho_j$ and $f_j$ have been radially interpolated 
using equation
(\ref{fdint}). Along with $w_1$, also the effective gravitational mass $M$ has
been linearly extrapolated in time at level $t^{n+1}$. At this point
$u_j^{n+1}$ can be interpolated from $u_j^{n+1/2}$ and $u_j^{n+3/2}$
and the value of $\Gamma_j^{n+1}$ can be updated.
The inner and outer boundary conditions for $u$ are (equations [\ref{ubc1}] and
[\ref{ubc2}])
\begin{eqnarray}
& & u_{j_{min}}^{n+3/2} = u_{j_{min}}^{n+1/2} + \Delta t^{n+1} 
a_{j_{min}}^{n+1}
\frac{ M_{j_{min}}^{n+1} }{ (r_{j_{min}}^{n+1})^2 }
\label{fd6bc1} \\
& & u_{j_{max}}^{n+3/2} = u_{j_{max}}^{n+1/2} \, ,
\label{fd6bc2}
\end{eqnarray}
where $a_{j_{min}}^{n+1}$ has been radially extrapolated from the neighboring
mesh points.

The first moment of the specific intensity is computed from equation
(\ref{mom1c}). In finite difference form it reads ($j_{min}+1 \leq j \leq 
j_{max}-1$)
\begin{equation}
(w_1)_j^{n+3/2} = \frac{1}{1 + F_w/2} \left[ \left( 1 - \frac{F_w}{2} \right)
(w_1)_j^{n+1/2} + H_w  \right] \, ,
\label{fd7}
\end{equation}
where
\begin{eqnarray}
F_w & = & \Delta t^{n+1} a_j^{n+1} \left[ (k_R)_j \rho_j + \frac{2}{r_j}
\frac{u_{j+1/2}r_{j+1/2} - u_{j-1/2}r_{j-1/2}}{r_{j+1/2} - r_{j-1/2}} 
\right]^{n+1} \\
H_w & = & \Delta t^{n+1} \left\{ \frac{8\pi a_j r_j}{\Gamma_j} (w_1)_j^2 -
a_j \Gamma_j \left[ \frac{ \left( 1/3 + f_{j+1/2} \right) (w_0)_{j+1/2}
- \left( 1/3 + f_{j-1/2} \right) (w_0)_{j-1/2} }{ r_{j+1/2} - r_{j-1/2} }
\right] \right. \nonumber \\
& & \left. - \Gamma_j \left( \frac{4}{3} + f_j \right) (w_0)_j
\frac{a_{j+1/2} - a_{j-1/2}}{r_{j+1/2} - r_{j-1/2}}
- 3 \frac{a_j \Gamma_j}{r_j} f_j (w_0)_j \right\}^{n+1} \, .
\label{fd7h}
\end{eqnarray}
As in the 0--th moment equation, to derive equation (\ref{fd7})
we have expressed $r_{,t}$ and $b_{,t}/b = - (\rho r^2)_{,t}/\rho r^2$
using equations (\ref{u}) and (\ref{continuity}).
To evaluate the radiation self--gravity term (the first term in curly brackets
in equation [\ref{fd7h}]) an extrapolated value (in time) of $w_1$ is used. 
The inner boundary condition is a 5 point Lagrangian extrapolation in radius.
Following \markcite{Mihalas \& Mihalas 1984}Mihalas \& Mihalas (1984) and
\markcite{Shapiro 1996}Shapiro (1996), we impose the outer boundary condition
applying equation (\ref{mom1c}) across the half--zone from $j_{max}-1/2$ to
$j_{max}$ and substituting $(w_1)_{j_{max}} = g (w_0)_{j_{max}}$ for
$(w_0)_{j_{max}}$ in the gradients with respect to $\mu$.
The closure boundary factor $g$ is given by equation (\ref{closg}).

Finally, the last quantity to be computed is the gravitational mass $M$.
Integrating equation (\ref{mass}) along the time slice at constant $t$, we 
obtain
($j_{min}+1 \leq j \leq j_{max}$)
\begin{equation}
M_j^{n+1} = M_{j-1}^{n+1} +
\frac{ 4\pi (r^2)_{j-1/2}^{n+1} }{ r_j^{n+1} - r_{j-1}^{n+1} }
\left[ \rho_{j-1/2} \left( 1 + \epsilon_{j-1/2} \right) + (w_0)_{j-1/2} +
\frac{u_{j-1/2}}{\Gamma_{j-1/2}} (w_1)_{j-1/2} \right]^{n+1} \, .
\label{fd8}
\end{equation}
The boundary condition at $\mu = \mu_{j_{min}}$ fixes 
$M_{j_{min}} = M_0$ (equation [\ref{mbc}]; see main text for explanation).

The timestep is computed at each cycle before the integration of the Euler
equation according to
\begin{eqnarray}
& & \Delta t^{n+3/2} = \min \left[ \Delta t_\rho, \Delta t_B, \Delta t_c,
(1.2 \Delta t^{n+1/2}) \right] \\
& & \Delta t^{n+1} = \frac{1}{2} (\Delta t^{n+1/2} + \Delta t^{n+3/2})
\label{timestep}
\end{eqnarray}
where ($j_{min}+1 \leq j \leq j_{max}$)
\begin{eqnarray}
& & \Delta t_\rho = \min \left[
0.05 \frac{ \rho_{j-1/2}^{n+1} }{ \rho_{j-1/2}^{n+1} - \rho_{j-1/2}^n }
\Delta t^{n+1/2} \right] \nonumber \\
& & \Delta t_B = \min \left[
0.1 \frac{ B_{j-1/2}^{n+1} }{ B_{j-1/2}^{n+1} - B_{j-1/2}^n }
\Delta t^{n+1/2} \right] \nonumber \\
& & \Delta t_c = \min \left[
0.4 \frac{ \Delta \mu_{j-1/2} }{ 4\pi \rho_{j-1/2}^{n+1} (r^2)_{j-1/2}^{n+1}
a_{j-1/2}^{n+1} (v_c)_{j-1/2}^{n+1} } \right]  \qquad \qquad {\rm Courant \ 
condition} \, . \nonumber
\end{eqnarray}
Here $v_c = (f + 1/3)^{1/2}$ is the radiation characteristic speed (the 
fastest speed on the grid).

Artificial viscosity is inserted into the finite difference equations
adding a scalar stress $q_{j-1/2}$ to the gas pressure $p_{j-1/2}$
(\markcite{von Neumann \& Richtmyer 1950}von Neumann \& Richtmyer 1950):
\begin{eqnarray}
q_{j-1/2} & = & 2 \rho_{j-1/2} (u_j - u_{j-1})^2  \qquad \qquad
{\rm if} \ u_{j-1} - u_j > 0 \\
& = & 0  \qquad \qquad \qquad \qquad \qquad \qquad {\rm otherwise} \, .
\end{eqnarray}

%\clearpage
\newpage

\figcaption{Light curve for model I. The observed
luminosity $l = L/L_{Edd}$ is plotted versus time $t$ in units
of the initial accretion timescale $t_{a,0}$.
The {\it squares} are the points computed numerically, while the
{\it solid line} represents the analytic solution by Arnett (1980).
%Squares denote $l$ as obtained from the numerical solution;
%the solid line denotes the analytic solution by Arnett. 
The arrow marks the diffusion timescale $t_{diff}$. \label{fig1}}

\figcaption{Light curve for model II. The conventions and symbols
are the same as in Figure 1. \label{fig2}}

\figcaption{Evolution of the
accretion luminosity and accretion rate at a fixed
radius $r_0 = 0.3 r_{a,0}$, where $r_{a,0} = GM_{bh}/c_{s,0}^2$
is the initial accretion radius, for model II.
(a) Luminosity $l(r_0)$ versus time $t$.
(b) Accretion rate ${\dot m} = {\dot M}_0/{\dot M}_{Edd}$
versus time $t$. \label{fig3}}

\figcaption{Properties of the flow at selected times for model II.
(a) Radial component of the 4--velocity $\vert u \vert$
in units of the initial sound velocity $c_{s,0}$ versus radius $r$
in units of $r_{a,0}$. The dashed line denotes negative values.
(b) Density $\rho$ in units of the initial density $\rho_0$ versus radius.
(c) Gas temperature $T$ versus radius.
(d) Luminosity $l$ versus $r/r_{a,0}$. At the onset
of evolution the radiative flux is taken to be zero.
Curves are labeled by time in unit of $t_{a,0}$. \label{fig4}}

\figcaption{Light curve for model III. The conventions and symbols
are the same as in Figure 1. \label{fig5}}

\figcaption{Luminosity $l$ versus radius $r/r_{a,0}$
at selected times for model III. \label{fig6}} 

\figcaption{Computed light curve for model IVa. The observed
luminosity $l = L/L_{Edd}$ ({\it squares}) is plotted versus time $t$
in units of the initial accretion timescale $t_{a,0}$.
When recombination takes
place ($t/t_{a,0} \sim 42$) the internal energy is released
and the luminosity suddenly increases.
%The maximum in the light curve is due to this effect.
After recombination the luminosity falls off exponentially
until the emission is dominated by accretion (note the slow power law
decline in $l$ al late times). \label{fig7}}

\figcaption{Properties of the flow at selected times for model IVa.
(a) Gas temperature $T$ versus radius. Note the formation
of the recombination front ($t/t_{a,0} \sim 42$) and its
rapid inward motion. The front stalls as soon as it approaches
the innermost accreting region.
(b) Luminosity $l$ versus $r/r_{a,0}$. 
%(c) Marginally bound radius ({\it squares}) and radius of the
%recombination front ({\it circles}) as a function of $t/t_{a,0}$.
(c) Radius of the recombination front $r_{rec}$ as a function of $t/t_{a,0}$.
\label{fig8}}

\figcaption{Light curve, effective temperature and photospheric
temperature for model IVb. (a) Luminosity $l$ ({\it squares})
versus time.
(b) Effective temperature $T_{eff}$ ({\it squares}), as defined
by equation (48), and gas temperature at the photospheric radius 
$T_{ph}$ ({\it circles}), as defined by equation (49), versus time.
%(c) Effective temperature $T_{eff}$ ({\it squares}) and
%gas temperature at the photospheric radius $T_{ph}$ ({\it circles})
%versus $t/t_{a,0}$.
%Scales are logarithmics.  
%The dot-dashed line is an extrapolation of the behavior of $T_{eff}$
%at late times.
\label{fig9}}

\figcaption{Early evolution of model IVb.
(a) Radial component of the 4--velocity $\vert u \vert$
%in units of the initial sound velocity $c_{s,0}$
versus radius $r$. The dashed line denotes negative values.
Note the sign inversion of the velocity of the intermediate shells
caused by the radiative force.
(b) Luminosity $l$ versus radius. The strong transient
superEddington flux produced early in the evolution ($t \sim t_{a,0}$)
pushes outwards the marginally bound shells near the accretion
radius and then propagates outwards. \label{fig10}}

\figcaption{Accretion rate ${\dot m}(r_0)$ at a fixed radius
($r_0 = 0.1 r_{a,0}$) versus time for
model IVa ({\it squares}) and model IVb ({\it circles}). \label{fig11}} 

\figcaption{Mass flux $\dot {m}$ versus radius
at $t = 6 \times 10^3 t_{a,0}$ for model IVb. Note the constancy
of ${\dot m}$ with radius in the inner accreting region. \label{fig12}}

\figcaption{Tracks on the luminosity--accretion rate plane
($l$--${\dot m}$) for model IVa and IVb. The observed luminosity
$l$ is plotted as a function of the accretion rate ${\dot m}$
(evaluated in the inner accreting region).
{\it Squares} denote the computed points for model IVa, while
{\it circles} denote those of model IVb.
The {\it solid line} is the analytic expression of $l = l({\dot m})$
for stationary models derived by Blondin (1986;
see equation (52) in the main text). \label{fig13}}

\figcaption{Computed and observed bolometric light curves
for SN1987A. The luminosity $L$ is plotted as a function of time $t$.
{\it Squares} ({\it circles}) give the solution for model IVa (IVb).
The {\it dotted lines} represent the extrapolation 
of the late-time evolution for the computed models (see Section 7).
%(see Section \ref{sec7}).
The {\it triangles} are the bolometric luminosity of SN1987A (data
taken by McCray 1993, Woosley \& Timmes 1996,
Arnett 1996 and Suntzeff 1997).
The {\it dashed lines} represent the expected contribution from the decay
of radioactive elements (0.07 $M_\odot$ of
$^{56}$Co and $\sim 5 \times 10^{-5} M_\odot$ of $^{44}$Ti).
Finally, the {\it upper dotted line} denotes the expected bolometric
luminosity emitted by a putative black hole in SN1987A. The {\it arrow}
marks the time ($t \simeq 900$ years)
at which the {\it upper dotted line}
crosses the dashed line (radioactive emission from $^{44}$Ti): 
from this moment on, the bolometric luminosity of SN1987A would
be dominated by the energy released by accretion. \label{fig14}}


\begin{references}

\reference{Arnett 1980} Arnett, D. 1980, \apj, 237, 541

\reference{Arnett 1996} Arnett, D. 1996, Supernovae and Nucleosynthesis
(Princeton: Princeton University Press)

\reference{Arnett et al. 1989} Arnett, D., Bahcall, J.N., Kirshner, R.P.,
\& Woosley, S.E. 1989, \araa, 27, 629

\reference{Berger \& Oliver 1984} Berger, M.J., \& Oliger, J. 1984,
Journal of Computational Physics, 53, 484

\reference{Blondin 1986} Blondin, J.M. 1986, \apj, 308, 755

\reference{Brown \& Bethe 1994} Brown, G.E., \& Bethe, H.A. 1994, \apj, 423, 659

\reference{Burrows 1988} Burrows, A. 1988, \apj, 334, 891

\reference{Chen \& Colgate 1995} Chen, K., \& Colgate, S.A. 1995, \apj, 
submitted

\reference{Chernoff, Shapiro and Wasserman 1989}Chernoff, D.F.,
Shapiro, S.L., \& Wasserman, I. 1989, \apj, 337, 814

\reference{Chevalier 1989} Chevalier, R.A. 1989, \apj, 346, 847

\reference{Chevalier 1996} Chevalier, R.A. 1996, \apj, 459, 322

\reference{Christy 1966} Christy, R.F. 1966, \apj, 144, 108

\reference{Colgate 1971} Colgate, S.A. 1971, \apj, 163, 221

\reference{Colgate 1988} Colgate, S.A. 1988, in Supernova 1987 A in the Large
Magellanic Cloud, ed. M. Kafatos and G. Michalitsianos (Cambridge:
Cambridge University Press), 341

\reference{CSW} Colpi, M., Shapiro, S.L., \& Wasserman, I. 1996, \apj, 470, 1075 
(CSW)

\reference{Cordes \& Chernoff 1997} Cordes, J., \& Chernoff, D. 1997, \apj, in press

\reference{Dar 1997} Dar, A. 1997, in the Proceedings of the 1997 Rencontre de
Physique de la Vallee d'Aoste, in press

\reference{Hansen \& Phinney 1997} Hansen, B.M.S., \& Sterl Phinney, E.
1997, \mnras, in press

\reference{Houck \& Chevalier 1991} Houck, J.C., \& Chevalier, R.A. 1991,
\apj, 376, 234

\reference{Hummer \& Rybicki 1971} Hummer, D.G., \& Rybicki, G.B. 1971,
\mnras, 152, 1

\reference{Kunkel et al. 1987} Kunkel, W., Madore, B., Shelton, I.K., 
Duhalde, O.,
Bateson, F. M., Jones, A., Moreno, B., Walker, S., McNaught, R.H.,
Garradd, G., Warner, B., \& Menzies, J. 1987, IAU Circular No. 4316

\reference{Lattimer \& Swesty 1991}
Lattimer, J.M., \& Swesty, F.D. 1991, Nucl. Phys. A, 535, 331

\reference{May \& White 1967} May, M.M., \& White, R.H. 1967,
in Methods in Computational Physics, Vol. 7 (New York: Academic Press), 219

\reference{McCray 1993} McCray, R. 1993, \araa, 31, 175

\reference{McCray, Shull and Sutherland 1987}McCray, R., Shull, J.M.,
\& Sutherland, P. 1987, \apj, 317, L73

\reference{Mihalas \& Mihalas 1984} Mihalas, D., \& Weibel Mihalas, B. 1984,
Foundations of Radiation Hydrodynamics (Oxford: Oxford University Press)

\reference{Nobili, Turolla \& Zampieri 1991}
Nobili, L., Turolla, R., \& Zampieri, L. 1991, \apj, 383, 250

\reference{Pandharipande 1997} Pandharipande, V.R. 1997, in the 
Proceedings of the Xth International Conference on Recent Progress in
Many-Body Theories, ed. D. Nelson (Singapore: World Scientific Publishing),
in press

\reference{Park 1990} Park, M.--G. 1990, \apj, 354, 64

\reference{Rezzolla \& Miller 1994} Rezzolla, L., \& Miller, J.C. 1994,
Class. Quantum Grav., 11, 1815

\reference{Rybicki \& Lightman 1979}
Rybicki, G.B., \& Lightman, A.P. 1979, Radiative Processes
in Astrophysics (New York: Wiley)

\reference{Shapiro 1996} Shapiro, S.L. 1996, \apj, 472, 308

\reference{Suntzeff 1997} Suntzeff, N.B. 1997, in SN1987A: Ten Years After, ed. 
M.M. Phillips and N.B. Suntzeff (ASP Conference Series), in press

\reference{Thorsson, Prakash \& Lattimer 1994}
Thorsson, V., Prakash, M., \& Lattimer, J.M. 1994, Nucl. Phys. A, 572, 693

\reference{von Neumann \& Richtmyer 1950}
von Neumann, J., \& Richtmyer, R.D. 1950, J. Appl. Phys., 21, 232

\reference{Wilson et al. 1986}
Wilson, J.R., Mayle, R., Woosley, S.E., \& Weaver T.A. 1986,
in Proceedings of the 12th Texas Symposium on Relativistic Astrophysics,
ed. M. Livio \& G. Shaviv (New York: New York Academy of Sciences), 267

\reference{Woosley 1988} Woosley, S.E. 1988, \apj, 330, 218

\reference{Woosley \& Weaver 1986} Woosley, S.E., \& Weaver T.A. 1986,
\araa, 24, 205

\reference{Woosley \& Weaver 1995} Woosley, S.E., \& Weaver T.A. 1995,
\apjs, 101, 181

\reference{Woosley \& Timmes 1996} Woosley, S.E., \& Timmes, F.X. 1996,
Nucl.Phys. A, 606, 137

\reference{Zampieri 1995} Zampieri, L. 1995, PhD thesis

\reference{Zampieri 1997} Zampieri, L. 1997, in the Proceedings of the Second
International Sakharov Conference on Physics, ed. I.M. Dremin \&
A.M. Semikhatov (Singapore: World Scientific Publishing), in press

\reference{Zampieri, Miller \& Turolla 1996}
Zampieri, L., Miller, J.C., \& Turolla, R. 1996, \mnras, 281, 1183

\end{references}
\end{document}